\documentclass[twocolumn,english,aps,prb]{revtex4}
\usepackage[T1]{fontenc}
\usepackage[latin9]{inputenc}
\usepackage{color}
\usepackage{amsmath}
\usepackage{graphicx}
\usepackage{epsf}

\newcommand{\nio}{Na$_2$IrO$_3$}
\newcommand{\lio}{Li$_2$IrO$_3$}
\newcommand{\aio}{A$_2$IrO$_3$}

\newcommand{\w}{\omega}
\newcommand{\e}{\epsilon}

\newcommand{\ket}[1]{\left| #1 \right\rangle}
\newcommand{\bra}[1]{\left\langle #1 \right|}

\newcommand{\HK}{\mathcal{H}_{\rm K}}
\newcommand{\HM}{\mathcal{H}_{\rm u}}

\newcommand{\HMh}{\mathcal{H}_{\rm \hat{u}}}

\newcommand{\Hz}{\mathcal{H}_{0}}

\newcommand{\mz}{M_0}
\newcommand{\mzp}{M_0^{\rm p}}
\newcommand{\ezp}{E_0^{\rm p}}

\begin{document}

\title{
Physical states and finite-size effects
in Kitaev's honeycomb model: \\
Bond disorder, spin excitations, and NMR lineshape
}

\author{Fabian Zschocke}
\author{Matthias Vojta}
\affiliation{Institut für Theoretische Physik, Technische Universit\"at Dresden,
01062 Dresden, Germany}


\begin{abstract}
Kitaev's compass model on the honeycomb lattice realizes a spin liquid whose emergent excitations are dispersive Majorana fermions and static Z$_2$ gauge fluxes. We discuss the proper selection of physical states for finite-size simulations in the Majorana representation, based on a recent paper by Pedrocchi, Chesi, and Loss [\prb {\bf 84}, 165414 (2011)]. Certain physical observables acquire large finite-size effects, in particular if the ground state is {\em not} fermion-free, which we prove to generally apply to the system in the gapless phase and with periodic boundary conditions.
To illustrate our findings, we compute the static and dynamic spin susceptibilities for finite-size systems. Specifically, we consider random-bond disorder (which preserves the solubility of the model), calculate the distribution of local flux gaps, and extract the NMR lineshape.
We also predict a transition to a random-flux state with increasing disorder.
\end{abstract}

\date{July 3, 2015}

\pacs{}

\maketitle

\section{Introduction}

Frustrated magnetism is an exciting field of research in condensed matter physics. Particular attention has been devoted to so-called spin-liquid states:\cite{balents10} In a stringent definition, these are zero-temperature states of local-moment systems with half-odd-integer spin per crystallographic unit cell which are characterized by the {\em absence} of any spontaneous symmetry breaking. Typically, the low-energy description of such states involves non-trivial elementary excitations with fractional quantum numbers which are coupled to an emergent gauge field.

In a seminal paper,\cite{kitaev06} Kitaev proposed a model of quantum spins $1/2$ on a two-dimensional
honeycomb lattice, subject to a particular type of anisotropic exchange interactions, often dubbed ``compass'' interactions.\cite{compass_rmp} This model is exactly solvable, thanks to an infinite set of conserved quantities. It realizes a non-trivial spin-liquid state which, depending on the interaction parameters, can be gapless or gapped. Its elementary excitations are dispersing spinless ``matter'' fermions which are coupled to a frozen Z$_2$ gauge field. ,
By now, many properties of the Kitaev model have been studied, including static\cite{kitaev06} and dynamic\cite{shank,knolle} spin correlations as well as the physics of isolated defects.\cite{willans10,willans11,dhochak10} In addition, variants of the Kitaev model on other lattices, both in two\cite{yao07,baskaran09,kiv1,kamfor10,kells11} and three\cite{siyu08,mandal09,trebst14} space dimensions, have been discussed.
In all cases, the most popular analytical treatment of the compass interactions utilizes a Majorana representation of spins. Subtleties in dealing with the corresponding enlarged Hilbert space have been pointed out.\cite{kitaev06,yao07,loss}

On the materials side, oxides of the family {\aio}, with magnetic iridium ions subject to strong spin-orbit coupling, have been proposed\cite{Cha10} to realize an exchange Hamiltonian of Kitaev type, supplemented by additional spin-symmetric Heisenberg interactions. The resulting Heisenberg-Kitaev model has been investigated extensively:\cite{Cha10,Jia11,Reu11,Bha12,Cha13,perkins12,eamv14} While the spin liquid is stable to small admixtures of Heisenberg interactions, larger perturbations destroy it in favor of a variety of magnetically ordered phases. Experimentally, both {\nio} and {\lio} have been found to display magnetic order at low temperatures,\cite{Sin10,Liu11,Sin12} and it has been speculated that pressure might be used to tune them towards the spin-liquid regime. However, the precise microscopic Hamiltonian describing the magnetism in {\aio} is under debate.\cite{Choi12,Maz12,Cha13,kee14,kimchi14,perkins14,rachel14}

In this paper, we consider the honeycomb-lattice Kitaev model with random-magnitude exchange interactions, i.e., bond randomness. The model remains exactly solvable and thus belongs to the rare cases of exactly solvable random spin models in dimensions $d\geq 2$. (Brief discussions of disorder in the Kitaev model have been given in Refs.~\onlinecite{willans11} and \onlinecite{lahtinen14}, and a Kitaev-style chiral spin-liquid model with random exchange was considered in Ref.~\onlinecite{fiete11}.)
We shall utilize the Majorana-fermion representation to investigate the magnetic response of the bond-disordered Kitaev spin liquid, in particular the NMR lineshape. Disorder is treated exactly via finite-size exact diagonalization.

Particular attention is paid to the proper selection of physical states in the Majorana representation,\cite{loss} which results in a condition on the parity of matter fermion excitations. While this condition generally depends on both the flux configuration and the system geometry, we are able to prove that, for clean systems with periodic boundary conditions and interaction parameters in the gapless phase, this parity must always be odd in the flux-free sector. Hence, the physical ground state is {\em not} fermion-free, but contains one matter fermion excitation. As we will show, this implies large finite-size effects for many observables.
As an aside, we point out that the ground state of the clean Kitaev model for certain small systems is {\em not} in the flux-free sector of the Z$_2$ gauge field. For large systems, we predict a quantum phase transition, upon increasing bond randomness, from a flux-free to a random-flux ground state.

The body of the paper is organized as follows:
In Section~\ref{sec:model} we introduce the random-bond Kitaev model together with its Majorana representation and the numerical solution in terms of free canonical fermions. The required projection to the physical Hilbert space is subject of Section~\ref{sec:phys}. Section~\ref{sec:susc} outlines the numerical calculation of the susceptibility. In Section~\ref{sec:resclean} we briefly show numerical results for observables in the clean system, with focus on their finite-size behavior. General aspects of quenched bond disorder in the Kitaev model are discussed in Section~\ref{sec:gen}, while concrete numerical results are presented in Section~\ref{sec:resdis}. The transition to the random-flux state is discussed in Section~\ref{sec:trans}. A summary closes the paper.
Technical aspects of the physical-state selection are relegated to the appendix, as is the comparison of the Majorana and exact solutions for a small system of four unit cells.


\section{Model and Majorana representation}
\label{sec:model}

\subsection{Random-bond Kitaev model}

\begin{figure}[t!]
\centering
\includegraphics[width=0.24\textwidth]{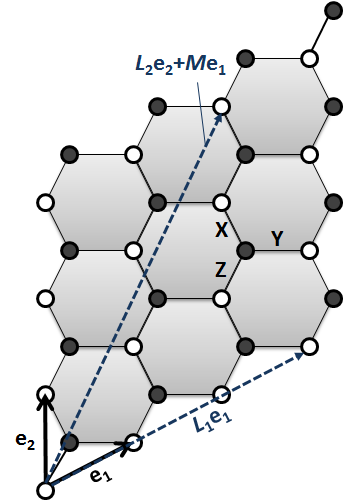}
\caption{
Honeycomb lattice with basis vectors ${\bf e}_{1,2}$ and an illustration of the periodic boundary conditions, characterized by the cluster size $L_{1,2}$ and the twist parameter $M$. The figure corresponds to $L_1=L_2=3$ and $M=2$.
}
\label{fig:BC}
\end{figure}

The Kitaev model\cite{kitaev06} describes spin-1/2 degrees of freedom at sites $i$ of a honeycomb
lattice which interact via Ising-like nearest-neighbor exchange interactions $J^\alpha$. The
anisotropy direction in spin space, $\alpha=x,y,z$, is coupled to the bond direction
in real space, reflecting a strong spin anisotropy from spin-orbit coupling.
We generalize the model to spatially varying, i.e., random, couplings, such that the
Hamiltonian reads
\begin{equation}
\label{hk}
\HK =
-\sum_{\langle ij\rangle_x} J_{ij}^x \hat{\sigma}_i^x \hat{\sigma}_j^x
-\sum_{\langle ij\rangle_y} J_{ij}^y \hat{\sigma}_i^y \hat{\sigma}_j^y
-\sum_{\langle ij\rangle_z} J_{ij}^z \hat{\sigma}_i^z \hat{\sigma}_j^z
\end{equation}
where $\hat{\sigma}_j^{\alpha}$ are Pauli matrices, and $\langle ij \rangle_\alpha$ denotes an $\alpha=x,y,z$ bond as in Fig.~\ref{fig:BC}.
In the clean case $J_{ij}^x=J^x$, $J_{ij}^y=J^y$, $J_{ij}^z=J^z$.
For isotropic couplings, $J^x=J^y=J^z\equiv J$, the model possesses a $Z_3$ symmetry of
combined real-space and spin rotations.

In our simulations of bond disorder, the exchange couplings $J_{ij}^\alpha$ will be drawn
from uncorrelated box distributions with mean value $J^\alpha>0$, $J_{ij}^\alpha \in
[J^\alpha-\Delta^\alpha,J^\alpha+\Delta^\alpha]$.
In a possible experimental realization in an insulating solid, disorder in the $J_{ij}$ arises from random lattice distortions and/or chemical disorder on non-magnetic sites, both of which locally modify individual exchange paths.

\subsection{Majorana representation}
\label{subsec:Mr}

Following Kitaev's solution,\cite{kitaev06} we introduce four (real) Majorana fermions $\hat{b}^x$, $\hat{b}^y$, $\hat{b}^z$ and $\hat{c}$.
Defining $\hat{\sigma}_i^\alpha=i \hat{b}_i^\alpha \hat{c}_i$, the original Hamiltonian in Eq.~\eqref{hk} can be mapped to
\begin{equation}
\label{hm}
\HMh = i \sum_{\langle ij\rangle}J^\alpha_{{ij}} \hat{u}_{ij}\hat{c}_i \hat{c}_j,
\end{equation}
where $\hat{u}_{ij}\equiv i\hat{b}_i^{\alpha_{ij}}\hat{b}_j^{\alpha_{ij}}$,
$\hat{u}_{ij}=-\hat{u}_{ji}$, and the summation is over all nearest-neighbor bonds. We follow the convention that, when specifying $\hat{u}_{ij}$, $i$ is located on sublattice $A$. The operators $\hat{u}_{ij}$, with
eigenvalues $u_{ij}=\pm 1$, commute with each other and with the Hamiltonian $\HM$, i.e.,
the $\{u_{ij}\}$ are constants of motion. A given set $\{u_{ij}\}$ reduces the
Hamiltonian to a bilinear in the $\hat{c}$ Majorana operators:
\begin{equation}
\label{hmflux}
\HM = \frac{i}{2}\left(\hat{c}^T_A \, \hat{c}^T_B\right) \begin{pmatrix}
0 & M \\
-M^T & 0
\end{pmatrix}
\begin{pmatrix}
\hat{c}_A \\
\hat{c}_B
\end{pmatrix}.
\end{equation}
Here $M$ is an $N \times N$ matrix with elements $M_{ij}=J^\alpha_{ij} u_{ij}$, and $\hat{c}_{A(B)}$ is a vector of $N$ Majorana operators on the $A(B)$ sublattice. Hence the problem takes the form of non-interacting Majorana fermions coupled to a static Z$_2$ gauge field.

The eigenmodes of $\HM$ can be found via singular-value decomposition of $M$, $M=USV^T$, where $U$ and $V$ are $N\times N$ orthogonal matrices, and $S$ is an $N\times N$ diagonal matrix containing the non-negative singular values of $M$. We define new Majorana operators according to
\begin{equation}
\begin{split}
\label{uvtrans}
(\hat{b}'_{1},\ldots,\hat{b}'_{N})&=(\hat{c}_{A,1},\ldots,\hat{c}_{A,N}) U \,,\\
(\hat{b}''_{1},\ldots,\hat{b}''_{N})&=(\hat{c}_{B,1},\ldots, \hat{c}_{B,N}) V \,.
\end{split}
\end{equation}
We may combine the transformation matrices $U$ and $V$ into a matrix $Q^u$,
\begin{equation}
\label{traf}
Q^u= \begin{pmatrix}
0 & U \\
V & 0
\end{pmatrix},
\end{equation}
which is equivalent to $Q^u$ defined in Eq.~(4) of Ref.~\onlinecite{loss} after re-ordering of both rows and columns.

For a given set of $\{u_{ij}\}$ the Hamiltonian now has the form $\HM = i\sum^N_{m=1} \e_m \hat{b}'_{m}\hat{b}''_{m}$, where $\e_m \geq 0$ are the singular values of $M$.
It is convenient to combine the Majorana operators $\hat{b}'$, $\hat{b}''$ into canonical fermions according to
\begin{equation}
\label{adef}
\hat{a}_m=\frac{1}{2}(\hat{b}'_{m} + i\hat{b}''_{m})\,.
\end{equation}
This eventually gives
\begin{equation}
\label{hmsvd}
\HM = \sum^N_{m=1} \e_m ( 2 \hat{a}^\dagger_{m}\hat{a}_{m}-1)
\end{equation}
with the ground-state energy $E_0=-\sum_m \e_m$.
Eigenstates of the Hamiltonian \eqref{hm} can thus be understood as a direct product of ``gauge'' ($u$) and ``matter'' ($a$) degrees of freedom.

\subsection{Boundary conditions}

To avoid edge effects, the analytical discussion as well as the numerical calculations will be performed for finite-size systems with periodic boundary conditions. We will comment on open boundary conditions in Section~\ref{sec:openbc} below.

As in Ref.~\onlinecite{loss}, we will restrict our attention to ``rectangular'' clusters of size $N=L_1 \times L_2$ unit cells, with $2N$ spins, but allow for a geometric ``twist'' characterized by an integer $M$ when imposing periodicity. Here, the torus is defined through the basis vectors $L_1{\bf e}_1$ and $L_2{\bf e}_2 + M{\bf e}_1$, see Fig.~\ref{fig:BC}. In the isotropic case this represents the most general set of periodic boundary conditions for rectangular clusters.

\subsection{Flux degrees of freedom}

For every closed loop $C$ of the lattice, the Kitaev model \eqref{hk} features a conserved quantity  $\hat{W}_C$.\cite{kitaev06,halasz} For a loop $C$ containing $L$ sites labeled $\{1,2,...,L\}$, the corresponding operator is
\begin{equation}
\hat{W}_C=\hat{\sigma}_1^{\alpha_{1,2}}\hat{\sigma}_2^{\alpha_{1,2}}\hat{\sigma}_2^{\alpha_{2,3}}\hat{\sigma}_3^{\alpha_{2,3}}\dots \hat{\sigma}_L^{\alpha_{L,1}} \hat{\sigma}_1^{\alpha_{L,1}}
\end{equation}
where $\alpha_{i,j}=x,y,$ or $z$ corresponds to the type of the bond connecting sites $i$ and $j$.
The eigenvalues of the $\hat{W}_C$ are $W_C=\pm1$, each corresponding to a $Z_2$ flux. It is convenient to introduce loop operators for the flux through each elementary plaquette of the lattice,
\begin{equation}
 \hat{W}_p = \hat{\sigma}_1^x \hat{\sigma}_2^y \hat{\sigma}_3^z \hat{\sigma}_4^x \hat{\sigma}_5^y \hat{\sigma}_6^z
\end{equation}
with $1, \ldots, 6$ labelling the sites of the plaquette under consideration.
For periodic boundary conditions, there are two additional (``topological'') loop operators $\hat{W}_{1,2}$ that wrap around the torus in the direction of the unit vectors ${\bf e}_{1,2}$ and are related to the flux through the torus holes.

A system with $N$ unit cells and periodic boundary conditions is characterized by $(N-1)$
independent plaquette fluxes $W_p$, due to the constraint\cite{kitaev06} $\prod_p W_p =
1$. Together with the torus fluxes $W_{1,2}$ the total number of flux degrees of freedom
is $(N+1)$.
Given that the dimension of the physical Hilbert space of $\HK$ is $2^{2N}$, this implies
that each individual flux sector consists of $2^{N-1}$ many-body states.

In the Majorana representation, the loop (or flux) operators $\hat{W}$ can be expressed through the bond variables $\hat{u}_{ij}$; the same holds for their eigenvalues. For instance, the plaquette fluxes take the form
\begin{equation}
W_p = u_{21}u_{23}u_{43}u_{45}u_{65}u_{61}\,.
\end{equation}
As a consequence of gauge invariance, the fermion spectrum $\{\e_m\}$ and the ground-state energy $E_0$ depend on the $u_{ij}$ only through the values of the fluxes, $\{W_p\}$ and $W_{1,2}$.

For a translation-invariant system of sufficiently large size, the ground state is
located\cite{kitaev06} in the flux-free sector, corresponding to all $W=1$ (i.e. all
$u_{ij}=1$). In this sector, the excitation spectrum of the hopping Hamiltonian $\HM$ can
be found using a Fourier transformation. Depending on the anisotropy of the couplings, the system is either gapped or gapless, with the latter case including the isotropic point, $J^x=J^y=J^z$.
Here, the low-energy part of the spectrum consists of two Dirac cones similar to graphene.
It is worth noting that the ground state of certain small systems is {\em not} in the
flux-free sector; this will be further discussed in Section~\ref{sec:gsflux}.


\section{Physical many-body states in the Majorana description}
\label{sec:phys}

The Majorana representation of spins 1/2 is overcomplete: The total Hilbert space of $\HMh$ has $4^{2N}$ states, as compared to $2^{2N}$ states forming the Hilbert space of $\HK$.
First, the $2^{N+1}$ physical flux sectors are represented by $2^{3N}$ link variables $u_{ij}$, such that different configurations of $\{u_{ij}\}$ correspond to the same flux sector.
Second, within each flux sector there are $2^N$ states of the $c$ Majorana fermions, to be
compared to $2^{N-1}$ physical states. This implies that the possible fermion+flux states
can be grouped into ``physical'' and ``unphysical'' states,\cite{yao07,loss} with not all physical
fermion+flux states corresponding to different spin states of the model \eqref{hk}.

\subsection{Projection}

We first summarize the Majorana state projection as outlined in Refs.~\onlinecite{kitaev06,loss} and then present our extended results in the following subsections.

An eigenstate of $\HK$, $\ket{\xi}$, satisfies the condition $\hat{D}_j \ket{\xi} =
\ket{\xi}$, where $\hat{D}_j\equiv-i\hat{\sigma}_j^x\hat{\sigma}_j^y \hat{\sigma}_j^z=1$.
Written in terms of Majorana operators, we have $\hat{D}_j=\hat{b}_j^x\hat{b}_j^y\hat{b}_j^z\hat{c}_j$, with eigenvalues $\pm1$. Now, a {\em physical} eigenstate must satisfy $\hat{D}_j\ket{\xi}=+\ket{\xi}$ for all $j$. One can therefore define a projection $\mathcal{\hat{P}}$ to the physical subspace of the Majorana Hilbert space according to
\begin{equation}
\label{P}
\mathcal{\hat{P}} = \prod^{2N}_{j=1} \left(\frac{1+\hat{D}_j}{2}\right).
\end{equation}
In this subspace the original spin Hamiltonian, $\HK$, and Kitaev's Majorana Hamiltonian
for the honeycomb lattice, $\HMh$, are equivalent.
The operator $\hat{D}_j$ can be thought of as an Ising gauge transformation. Since the spin operators are gauge-invariant, their matrix elements in any gauge-fixed sector are identical to that in the physical gauge-invariant subspace.\cite{shank}

The effect of the projection (\ref{P}) to annihilate an unphysical state is easily seen by rewriting it as \cite{kiv1}
\begin{equation}
 \mathcal{\hat{P}}=\mathcal{\hat{S}}\left(\frac{1 +\prod_{j=1}^{2N} \hat{D}_j}{2}\right)=\mathcal{\hat{S}} \mathcal{\hat{P}}_0 ,
\end{equation}
where $\mathcal{\hat{S}}$ symmetrizes over all gauge-equivalent subspaces while $\mathcal{\hat{P}}_0$
projects out unphysical states.

The operator $\hat{D} = \prod_j \hat{D}_j$ can be expressed in the Majorana representation. After re-ordering the fermion operators -- see Appendix B of Ref.~\onlinecite{loss} -- it can be brought into the form:
\begin{equation}
\label{dmajorana}
\hat{D} = (-1)^\theta \prod_{j} \hat{c}_j \prod_{\left< ij\right>_\alpha} \hat{b}_i^\alpha \hat{b}_j^\alpha = (-1)^\theta \hat{\pi}_c  \prod_{\left< ij \right>} u_{ij}.
\end{equation}
Here, $\hat{\pi}_c=i^N\prod_j \hat{c}_j$ is the parity of the $c$ (matter) Majorana fermions,
and we followed the convention that sites labeled with odd (even) numbers belong to the $A (B)$ sublattice (this differs from Ref.~\onlinecite{loss}).
The exponent $\theta$ is a consequence of the anticommutation relation of the Majorana fermions and depends on the lattice geometry. For the boundary conditions in Fig.~\ref{fig:BC} it reads \cite{loss}
\begin{equation}
\label{eq:gf}
\theta = L_1 +L_2 +M(L_1-M).
\end{equation}

An alternative representation of the Majorana states uses local complex fermions. For each unit cell ${\bf r}$ one can construct one complex matter fermion
\begin{equation}
\hat{f}_{\bf r} = \frac{1}{2} \left[ \hat{c}_{A,r} - i \hat{c}_{B,r} \right]
\end{equation}
and three complex gauge fermions defined on the bonds emanating from site $i$ on sublattice $A$: \begin{equation}
\label{chidef}
\hat{\chi}_{\bf r}^{\alpha} = \frac{1}{2} \left[ \hat{b}_i^{\alpha_{ij}} - i \hat{b}_j^{\alpha_{ij}} \right].
\end{equation}
Then we have $ i \hat{c}_{A,r} \hat{c}_{B,r} = 1- 2 \hat{f}_{\bf r}^\dagger \hat{f}_{\bf r}$ such that we can express the parity $\pi_c$ as $\pi_c=(-1)^{N_f}$ with $N_f = \sum_{\bf r}  \hat{f}_{\bf r}^\dagger  \hat{f}_{\bf r}$.
Similarly, $i\hat{b}_i^{\alpha}\hat{b}_j^{\alpha} = \hat{u}_{ij} = 1-2\left( \hat{\chi}_{\bf r}^{\alpha}\right)^\dagger \hat{\chi}_{\bf r}^{\alpha}$ which yields $\prod_{\left< ij \right>} u_{ij}=(-1)^{N_\chi}$.
This allows one to rewrite the operator $\hat{D}$ \eqref{dmajorana} using the fermion numbers $N_f$ and $N_\chi$:
\begin{equation}
\label{dbond}
\hat{D} = (-1)^\theta (-1)^{N_f} (-1)^{N_\chi}.
\end{equation}
The condition for a state being physical, $\hat{D}=2\mathcal{\hat{P}}_0-1 \overset{!}{=}1$, selects states with either even or odd total fermion number, depending on the geometry factor $(-1)^\theta$. For fixed $\{u_{ij}\}$ this eliminates half of the many-body states from the Hilbert space of $\HM$, as anticipated, and implies that fermions can only be excited pairwise. We note that the factor $(-1)^\theta$, derived in Ref.~\onlinecite{loss}, does not seem to appear in earlier works.\cite{willansfoot}

To convert Eq.~\eqref{dbond} into a more useful form, it is important to distinguish the parity $\hat{\pi}_c$ of the $\hat{c}$ fermions from the parity $\hat{\pi}=\prod_m^N(1-2\hat{a}^\dagger_{m}\hat{a}_{m})$ of the eigenmodes $\hat{a}_m$ \eqref{adef}.
Given that the $\hat{c}$ and $\hat{a}$ fermions are related via the canonical transformation $Q^u$, Eq.~\eqref{traf}, one finds\cite{loss}
\begin{equation}
\label{parity}
 \hat{\pi}_c=\det(Q^u)\hat{\pi}\,.
\end{equation}
Combining Eqs.~\eqref{dbond} and \eqref{parity} the operator $\hat{D}$ reads
\begin{equation}
\label{Dmatter}
\hat{D} = (-1)^\theta \det(Q^u)(-1)^{N_a} (-1)^{N_\chi}
\end{equation}
with $N_a = \sum_m \hat{a}^\dagger_{m}\hat{a}_{m}$ being the number of matter fermion excitations.

\subsection{Fermion parity for periodic boundary conditions}
\label{sec:pbc}

In general, the value of $\hat{D}$ \eqref{Dmatter} depends in a combined fashion on the flux configuration, the boundary conditions, and the distribution of the coupling constants; concrete examples were given in Ref.~\onlinecite{loss}.

Here we go one step further:
For a translation-invariant system in the gapless phase, we are able to prove that in the flux-free sector we have $(-1)^\theta \det(Q^u) = -1$ {\em independent} on the system geometry. Details of this proof are given in Appendix~\ref{sec:physicalgapless}. Since the flux-free sector is characterized by $N_\chi =0$, the condition $\hat{D}\overset{!}{=}1$ translates into $\hat{\pi}=(-1)^{N_a}\overset{!}{=}-1$, i.e., all physical states in the flux-free sector must have an {\em odd} number of $\hat{a}$ fermion excitations. Hence, the naive fermion-free state is not a physical state. This has consequences for the calculation of observables, as will be discussed below.

On general grounds, we expect that a single fermion in an extended system of size $N$ can cause only $1/N$ effects on observables. Hence, the proper selection of physical states discussed here, albeit important for finite-size systems, is not expected to influence typical observables in the thermodynamic limit. Indeed, in our calculations we find strong differences in the finite-size behavior of observables calculated with either physical or unphysical states, but these differences diminish with increasing system size. However, for observables where $1/N$ corrections are crucial -- this applies to quantum impurity problems -- the state of affairs might be different; this will be investigated in future work.

\subsection{Fermion parity for open boundary conditions: dangling gauge fermions}
\label{sec:openbc}

The considerations in Ref.~\onlinecite{loss} and the present section show that, for a Kitaev model with periodic boundary conditions, half of the Majorana many-body states are unphysical. Formally, the unphysical states do not obey the condition on total fermion parity imposed by the projector.

Although not the main focus of this work, it is interesting to repeat the analysis with open boundary conditions. More generally, we may consider a lattice with formally periodic boundary conditions, but allow for an arbitrary number of ``missing'' bonds with zero bond strength $J_{ij}$; this includes the cases of both open and cylindric boundary conditions.

A missing $\alpha$ bond, connecting sites $i$ and $j$, induces two dangling gauge Majorana fermions, $b_i^\alpha$ and $b_j^\alpha$. These can be combined into a canonical fermion, Eq.~\eqref{chidef}, which is decoupled (for zero external field), hence represents a zero-energy mode. Occupying this zero mode obviously changes the total fermion parity without changing observable properties of the many-body state. As a result, a given Majorana many-body state can always be turned from physical to unphysical or vice versa by changing the zero-mode occupation. Phrased differently, all {\em matter} Majorana states in any flux sector are physical if there is at least one missing bond which can ``absorb'' the fermion-parity condition. In Appendix~\ref{app:exact} we demonstrate this for a small $2\times2$ system.
A consequence is that the number of fermion zero modes of a Kitaev model with missing bonds is smaller by {\em one} compared to the number of zero modes suggested by its Majorana representation.


\section{Spin correlations and magnetic susceptbility}
\label{sec:susc}

Dynamical spin correlations in the Kitaev model have been calculated in Ref.~\onlinecite{knolle}. In this section we summarize and extend the required formalism.

Consider the zero-temperature spin correlation function
\begin{equation}
S^{\alpha \beta}_{ij}(t)=\bra{0}\hat{\sigma}_i^\alpha(t)\hat{\sigma}_j^\beta(0)\ket{0}
\end{equation}
where $\ket{0}$ is the many-body ground state. Given that the fluxes are constants of motion, the correlator can be calculated by decomposing the ground state $\ket{0}$ as a direct product of the ground states in the gauge and matter sector.
Specifically, the application of a $\hat{\sigma}_i^\alpha$ operator changes the
two flux variables which involve the $\alpha$ bond emanating from site $i$. This leads to the dynamical rearrangement of matter fermions in the modified gauge field.
The spin correlator can therefore be expressed purely in terms of matter fermions in the ground-state flux sector, subject to a perturbation $\hat{V}_\alpha = -2iJ^\alpha c_i c_j$: \cite{shank,knolle}
\begin{equation}
\label{soft}
S^{\alpha \beta}_{ij}(t)=-i\bra{\mzp} e^{i\Hz t}\hat{c_i}e^{-i(\Hz+\hat{V}_\alpha)t} \hat{c}_{j}\ket{\mzp} \delta_{\alpha \beta} \delta_{\langle ij \rangle_\alpha}
\end{equation}
where $\Hz$ is the Majorana hopping Hamiltonian in the zero-flux sector and $\ket{\mzp}$ its {\em physical} ground state.
Site-off-diagonal contributions vanish beyond nearest neighbor pairs indicated by $\langle ij\rangle_a$. Site-diagonal terms are calculated similarly.
$\Hz + \hat{V}_\alpha$ and $\Hz$ differ in the sign of the Majorana hopping on the
$\alpha$-bond, representing the insertion of the flux pair.
A suitable Lehmann representation of Eq.~\eqref{soft} is in terms of the matter Majorana eigenstates of the Hamiltonian $\Hz+ \hat{V}_\alpha$, denoted by $\ket{\lambda}$:
\begin{equation}
\label{sofw}
\begin{split}
S^{\alpha \beta}_{ij}(\omega)=&-i\sum_\lambda \bra{\mzp}\hat{c}_i\ket{\lambda} \bra{\lambda}\hat{c}_j\ket{\mzp} \\
& \times \delta[\omega-(E_\lambda-\ezp)] \delta_{\langle ij \rangle_\alpha} \delta_{\alpha \beta}.
\end{split}
\end{equation}
Here, $\ezp$ and $E_\lambda$ are the energies of the initial and intermediate states. In the following, the complete sum over excited states $\ket{\lambda}$ will be approximately evaluated using states with a fixed (small) number of matter excitations of $\Hz+ \hat{V}_\alpha$; this is a suitable strategy provided that no orthogonality catastrophe occurs.\cite{knolle}


In order to evaluate the matrix elements $\bra{\mzp}\hat{c}_i\ket{\lambda}$, involving eigenstates of both $\Hz+ \hat{V}_\alpha$ and $\Hz$, we need a conversion for the excitation operators. In the following we denote the operators for matter eigenmodes in the zero-flux and two-flux sectors with $\hat{a}$ and $\hat{b}$, respectively. As in Eq.~\eqref{uvtrans}, these are constructed from the matter Majorana operators according to
\begin{equation}
 \begin{split}
  (\hat{a}_1, \ldots, \hat{a}_N) &=\frac{1}{2} \left[(\hat{c}^T_A)U+i(\hat{c}^T_B)V\right], \\
  (\hat{b}_1, \ldots, \hat{b}_N) &=\frac{1}{2} \left[(\hat{c}^T_A)U'+i(\hat{c}^T_B)V'\right].
 \end{split}
\end{equation}
Using a Bogoliubov transformation, one can express the one kind of operators in terms of the other
\begin{equation}
 \hat{b}_\lambda = \sum_{m} X^{*}_{\lambda {m}} \hat{a}_{m} + Y^{*}_{\lambda m} \hat{a}^\dagger_{m}
\end{equation}
where $X,Y$ are the transformation matrices
\begin{equation}
 \begin{split}
  X^{*} &=\frac{1}{2}(U'^\dagger U+V'^\dagger V),\\
  Y^{*} &=\frac{1}{2}(U'^\dagger U-V'^\dagger V)
 \end{split}
\end{equation}
which obey the conditions \cite{BlaRip}
\begin{equation}
 \begin{split}
  XX^{\dagger} + YY^{\dagger} &=1, \quad XY^T + YX^{T} =0, \\
  X^{\dagger}X+Y^TY^{*} &=1,      \quad  X^TY^{*}+Y^{\dagger}X =0.
 \end{split}
\end{equation}
This allows one to rewrite the fermion-free state of the two-flux sector, $\ket{\lambda_0}$, in terms of $\hat{a}$ fermions and the fermion-free state in the zero-flux sector, $\ket{\mz}$:
\begin{equation}
 \ket{\lambda_0} = \left[X^\dagger X\right]^{1/4} e^{-\frac{1}{2} {\bf \hat{a}}^\dagger X^{* -1} Y^{*} {\bf \hat{a}}^\dagger }\ket{\mz},
\end{equation}
with the overlap $\left|\langle\mz|\lambda_0 \rangle \right|=\sqrt{\left|\det X\right|}$.\cite{knolle}

However, as we have pointed out in Section~\ref{sec:phys}, in the gapless phase the physical states in the flux-free sector must have an odd number of $\hat{a}$ fermions. Hence, $\ket{\mzp}=\hat{a}_1^\dagger\ket{\mz}$ and $\ezp = E_0^{(0)}+2\epsilon_1^{(0)}$ where $E_0^{(0)}$ and $\epsilon_1^{(0)}$ are the energies of the ground state and the lowest excitation of $\HM$ in the flux-free sector.
Using Eq.~\eqref{Dmatter} we find that $\ket{\lambda}$ must contain an even number of matter fermion excitations, see Appendix~\ref{sec:physicalgapless}.
These multi-particle eigenstates of $\Hz+ \hat{V}_\alpha$ are given by $\ket{\lambda}=\hat{b}_{\lambda_n}^\dagger \ldots \hat{b}_{\lambda_1}^\dagger \ket{\lambda_0}$,
with $n$ even.
The simplest contribution to $\bra{\mzp}\hat{c}_i\ket{\lambda}$ is the zero-particle contribution:
\begin{equation}
\label{zpart}
 \bra{\mz}\hat{a}_1 \hat{c}_{A,i}\ket{\lambda_0} = \sqrt{\left|\det X\right|} {\Big [} U_{i0}-\left(UX^{-1}Y\right)_{i0}{\Big ]}
\end{equation}
written for $i$ on the $A$ sublattice. The two-particle contributions can be obtained by straightforward algebra as
\begin{multline}
\label{tpart}
\bra{\mz} \hat{a}_1 \hat{c}_{A,i} \hat{b}_{\lambda_2}^\dagger \hat{b}_{\lambda_1}^\dagger \ket{\lambda_0} = \\
\sqrt{\left|\det X\right|} {\Big [} U_{i0} \left(YX^{-1}\right)_{\lambda_1\lambda_2}+
\left(UX^{-1}\right)_{i\lambda_1} X_{\lambda_2 0}-\\
\left(UX^{-1}\right)_{i\lambda_2} X_{\lambda_1 0} + \left(UX^{T}\right)_{i\lambda_1} \left[X^{-1}_{0 \lambda_2}-X_{\lambda_2 0}\right] - \\
\left(UX^{T}\right)_{i\lambda_2} \left[X^{-1}_{0 \lambda_1}-X_{\lambda_1 0}\right] -
\left( UX^{-1}Y\right)_{i 0} \left(YX^{T}\right)_{\lambda_1 \lambda_2} \\
\left(UX^{-1}Y\right)_{i0} \left(XY^T\right)_{\lambda_1 \lambda_2} + \\
Y_{\lambda_2 0} \left(UY^T\right)_{i\lambda_2} - Y_{\lambda_1 0} \left(UY^T\right)_{i\lambda_1} {\Big ]}.
\end{multline}
Matrix elements for $\hat{c}_{B,j}$ are calculated similarly.\cite{kn_raman}

In contrast, upon ignoring the fermion parity condition one may start with the fermion-free state $\ket{\mz}$ in the zero-flux sector. Then, the spin correlation function starts with the one-particle contribution:
\begin{equation}
\label{opart}
 \bra{\mz} \hat{c}_{A,i} \hat{b}_{\lambda_1}^\dagger \ket{\lambda_0} = \sqrt{\left|\det X\right|} \left(UX^{-1}\right)_{i\lambda}
\end{equation}
and the energy $E_0$ appearing in Eq.~\eqref{sofw} is given by $E_0^{(0)}$.

%

Below we will show results for the dynamic structure factor at momentum ${\bf q}=0$,
\begin{equation}
\label{sqz}
S^{\alpha\alpha}({\bf q}=0,\omega) = \sum_{ij} S^{\alpha \alpha}_{ij}(\omega)
\end{equation}
and the static susceptibility $\chi_{ij}$, obtained via the Kramers-Kronig relation
\begin{equation}
\chi_{ij}^{\alpha \beta}(\omega=0)=- \mathcal{P} \int \mathrm{d}\omega' \frac{S^{\alpha \beta}_{ij}(\omega')}{\omega-\omega'},
\end{equation}
where $\mathcal{P}$ denotes the Cauchy principal value.


\section{Numerical results: Clean system}
\label{sec:resclean}

Applying the methodology outlined so far, we now exhibit a few numerical results for the clean Kitaev model, obtained via singular-value decomposition of the matrix $M$ in Eq.~\eqref{hmflux}.
We have treated finite-size systems with $L_{1,2}\leq 150$.
Unless noted otherwise, the magnetic couplings are chosen to be isotropic, $J^x=J^y=J^z\equiv J$.

\subsection{Finite-size behavior of the flux gap}
\label{sec:fluxgap}

\begin{figure}[h!]
\centering
\includegraphics[width=0.44\textwidth]{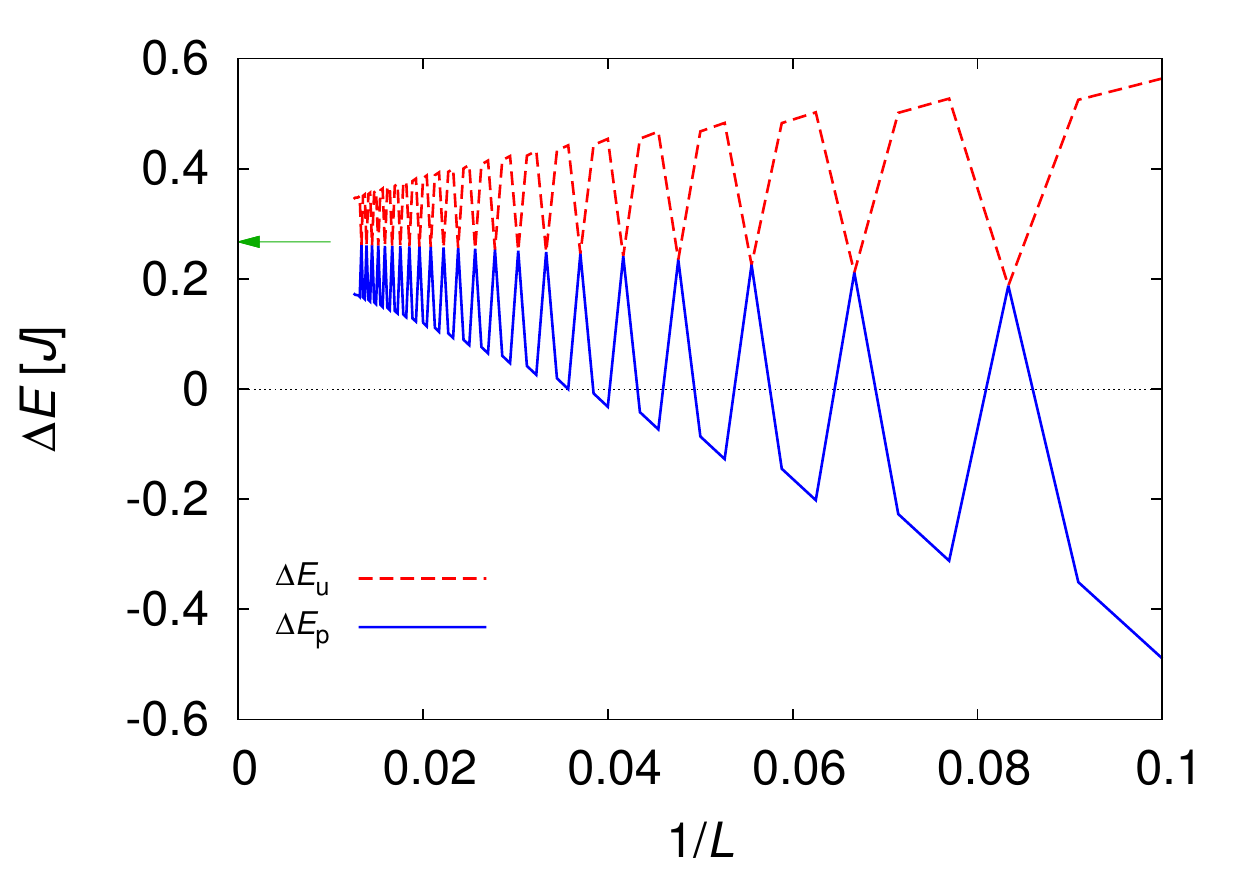}
\caption{
Flux gap $\Delta E$ of the isotropic Kitaev model as function of inverse system size, with $L_1=L_2\equiv L$, periodic boundary conditions, and $M=0$. The solid line shows the physical result, Eq.~\eqref{fluxgap1}, taking into account the presence of an excited matter fermion in the flux-free sector. In contrast, the dashed line shows the result \eqref{fluxgap2} where both the states in the flux-free and two-flux sectors are unphysical. $\Delta E_p = \Delta E_u$ is realized for $L \mod 3=0$ where the Dirac point is an allowed wavevector.
The arrow indicates the infinite-system result\cite{kitaev06} $\Delta E \approx 0.26 J$
}
\label{fig:gap}
\end{figure}

In Fig.~\ref{fig:gap} we show the finite-size scaling of the energy necessary to create a flux pair. Since the physical flux-free ground state contains one matter fermion excitation, whereas the lowest two-flux state does not, the physical energy gap is given by
\begin{equation}
\label{fluxgap1}
\Delta E_p = E_0^{(2)} - \ezp = E_0^{(2)}-(E_0^{(0)}+2\epsilon_1^{(0)})
\end{equation}
where $E_0^{(0)}$ and $E_0^{(2)}$ are the ground-state energies of $\HM$ in the zero-flux and two-flux sectors, respectively, and $\epsilon_1^{(0)}$ refers to the lowest singular value of $M$ in the flux-free sector. Alternatively, one may consider an unphysical gap,
\begin{equation}
\Delta E_u = (E^{(2)}_0+2\epsilon^{(2)}_{1}) -E_0^{(0)}
\label{fluxgap2}
\end{equation}
which involves states with incorrect fermion parity in both flux sectors.

As the $L=\infty$ matter fermion spectrum is gapless, we have $\epsilon^{(0)}_{1} = \epsilon^{(2)}_{1} = 0$ and thus $\Delta E_p = \Delta E_u$ whenever the Dirac point is included in the discrete set of momenta. For $M=0$ this applies to $L \mod 3=0$ -- these data points display weak $L$ dependence in Fig.~\ref{fig:gap}. In contrast, the data points for $L \mod 3 \neq 0$ are influenced by the strong $L$ dependence of $\epsilon^{(0)}_{1}$ or $\epsilon^{(2)}_{1}$.
We note that the result in Fig.~\ref{fig:gap} is qualitatively similar to that in Fig.~4 of Ref.~\onlinecite{loss} where different boundary conditions were employed.

Fig.~\ref{fig:gap} demonstrates that observables calculated for physical and unphysical states have rather different finite-size behavior; in particular the finite-size convergence appears significantly slower in the physical case. Knowing that both $\Delta E_p$ and $\Delta E_u$ have to converge to the same value as $L\to\infty$, one may choose the most suitable set of states and boundary conditions for fast convergence.

\subsection{Ground-state flux sector}
\label{sec:gsflux}

A remark is in order concerning negative values of the flux gap for small $L$, Fig.~\ref{fig:gap}, which imply that the ground state is {\em not} flux-free. It has been argued \cite{kitaev06} that a theorem of Lieb,\cite{lieb} being concerned with free-particle hopping Hamiltonians, guarantees that the ground state of the Kitaev model is always in the flux-free sector. This assertion is apparently incorrect, Fig.~\ref{fig:gap}, and the reasons are twofold:
(i) The theorem of Lieb applies to ground states of hopping Hamiltonians, but as established in Ref.~\onlinecite{loss} and here, the physical ground state of the Kitaev model may contain an excited matter fermion which changes the energetics (and in particular lowers the energy of the lowest many-body state in the two-flux sector relative to that in the flux-free sector).
(ii) Only systems with $L_2=M$ obey the particular periodicity requirement needed for Lieb's theorem to apply.
Taken together, the theorem of Lieb ensures that the ground state of the Kitaev model is in the flux-free sector in the limit of large system size (where the restrictions (i) and (ii) become irrelevant), but is not decisive for small systems.

\subsection{Finite-size behavior of the dynamic susceptibility}
\label{sec:fssus}

\begin{figure}[t!]
  \centering
    \includegraphics[width=0.45\textwidth]{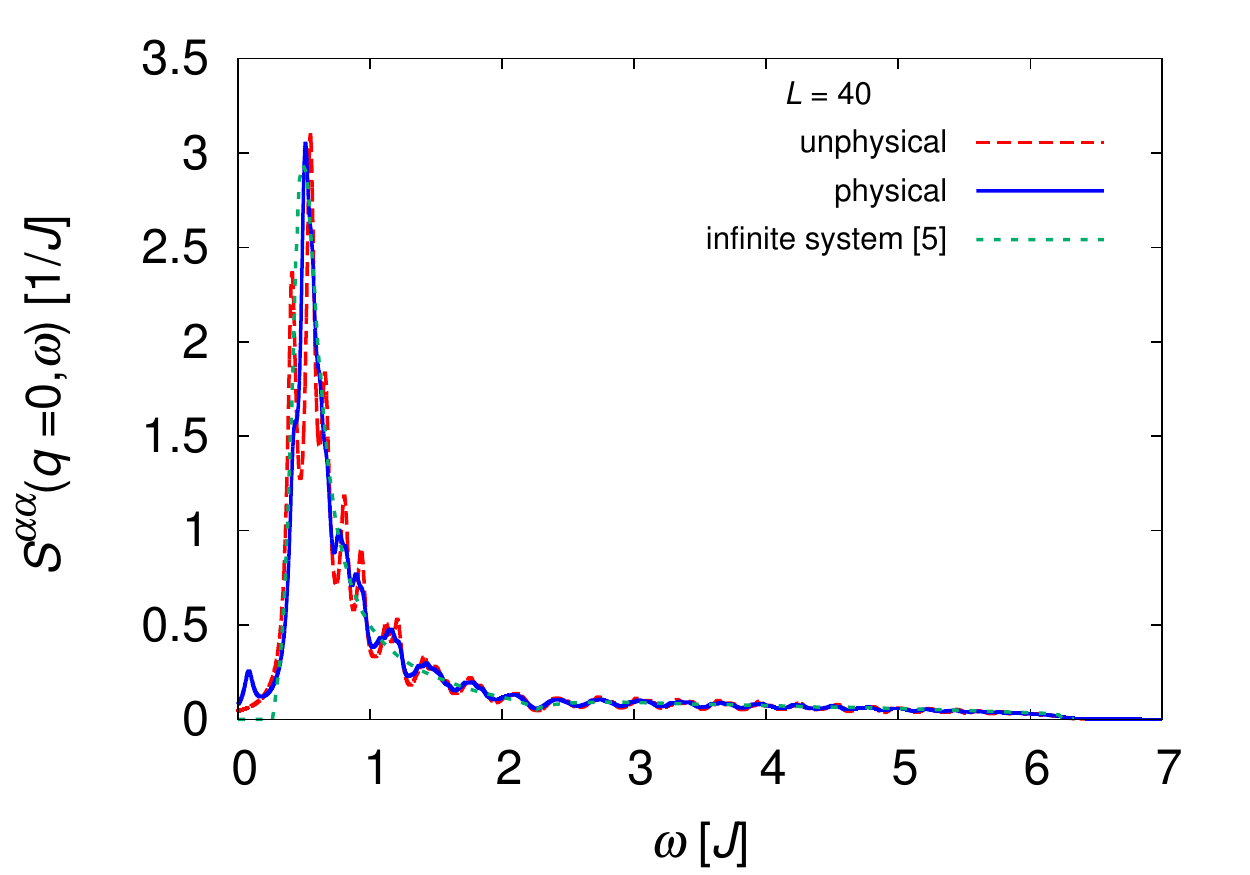}
    \includegraphics[width=0.45\textwidth]{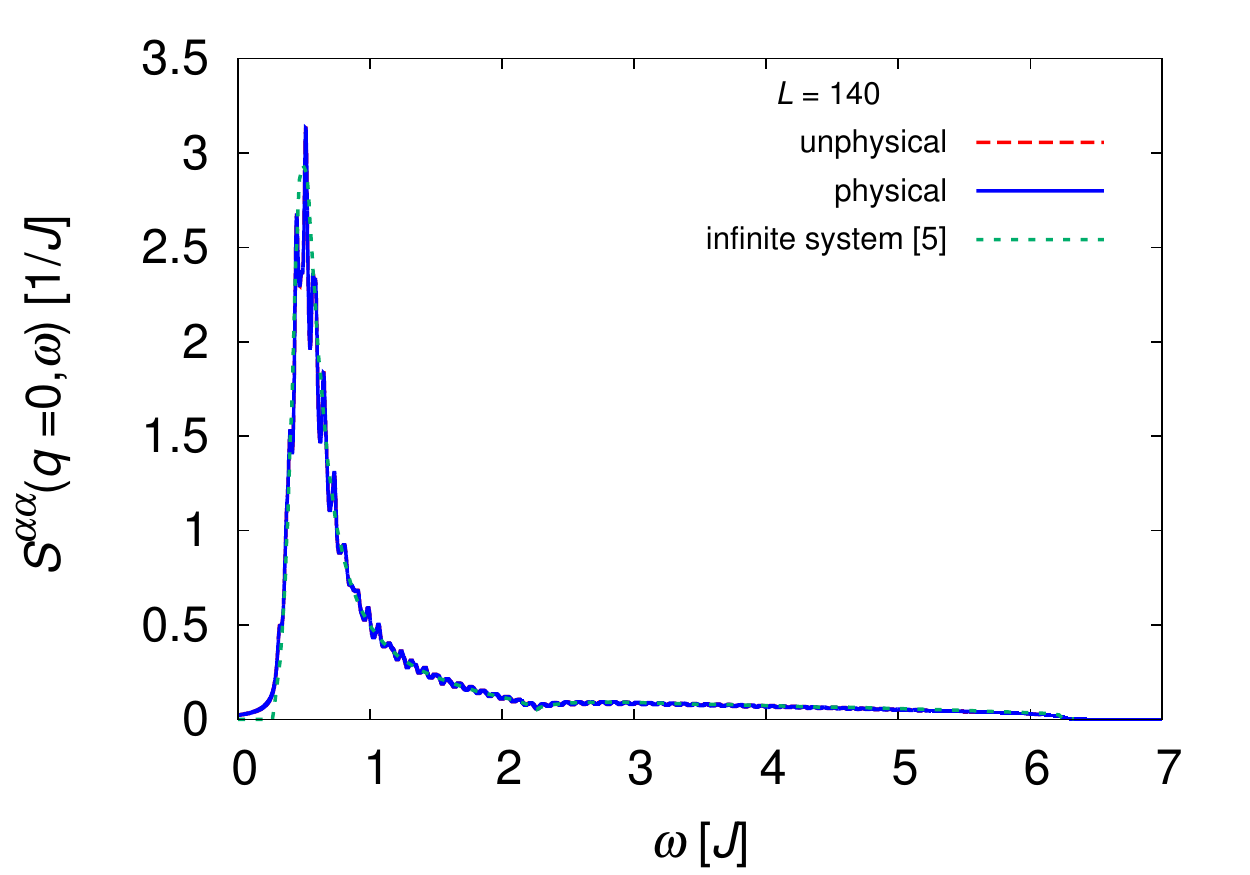}
\caption{
Dynamic structure factor for the isotropic Kitaev model, calculated from Eq.~\eqref{sqz} for systems with $L_{1,2}=40$ and a broadening of $\delta/J=0.04$ (top) and $L_{1,2}=140$ and $\delta/J=0.02$ (bottom), both with $M=0$.
The ``physical'' (solid) result takes into account the presence of a matter fermion in the ground state; it consists of two-particle contributions, Eq.~\eqref{tpart}, and an isolated low-energy peak corresponding to the zero-particle contribution, Eq.~\eqref{zpart}. In contrast, the ``unphysical'' result (dashed) contains one-particle contributions, Eq.~\eqref{opart}, only. The exact result \cite{knolle} for $L=\infty$ is shown for comparison.
}
\label{fig:s40}
\end{figure}

Fig.~\ref{fig:s40} shows the dynamical structure factor calculated for two system sizes. Reasonable finite-size convergence is apparent,\cite{orth_foot} and the results for $L=140$ are very close to the infinite-system result from Ref.~\onlinecite{knolle} -- the latter is known to have a gap of size $\Delta E/J \approx 0.26$, the flux gap.

Let us briefly discuss the difference between the physical and unphysical results. As explained in Section~\ref{sec:susc}, the physical flux-free ground state comes with one matter fermion excitation, such that (at the isotropic point) the excited intermediate states in the two-flux sector have an even number of matter fermions. In particular, there is a contribution from the zero-fermion intermediate state -- this produces an isolated $\delta$ peak in $S(\w)$ at low energies (clearly visible in the $L=40$ data at $\w/J \approx 0.08$). The rest of the signal comes from two-fermion intermediate states; higher excited states are ignored in our calculation because they only carry spectral weight of about 2.5\%.\cite{knolle}
In contrast, the unphysical signal is obtained by starting from a fermion-free ground state in the flux-free sector. Then, the signal at the isotropic point arises from single-fermion intermediate states, and the low-energy $\delta$ peak is absent.

Remarkably, the differences between the physical and unphysical signal diminish with increasing system size, in accordance with the general argument from Section~\ref{sec:pbc}. Here, the reason for this can be understood in detail:
Although the two-fermion intermediate states $\ket{\lambda}$ in the physical case can have two {\em arbitrary} fermions excited, the matrix element $\bra{\lambda} \hat{c}_i \ket{\mzp}$ will only be sizeable if one of the fermions is the lowest-energy one, simply to match the lowest-energy fermion occupied in $\ket{\mzp}$. All other matrix elements are suppressed at least with $N^{-1/2}$, which effectively reduces the two-particle continuum to the single-particle continuum of the unphysical case. Similarly, the matrix element $\bra{\lambda_0} \hat{c}_i \ket{\mzp}$, determining the weight on the low-energy $\delta$ peak in the physical response, scales as $N^{-1/2}$.
Hence, the dynamical structure factor in the thermodynamic limit is independent of the ground-state parity $\pi$.\cite{knollefoot}


\section{Bond disorder: General considerations}
\label{sec:gen}

Before showing numerical results for the Kitaev model with bond disorder, we quickly summarize a few general aspects, some of which have been discussed in Refs.~\onlinecite{willans10,willans11,fiete11}.

Provided that the ground state in the presence of disorder remains in the flux-free sector, the low-energy behavior in the presence of bond disorder is equivalent to that of Dirac fermions with random hopping on the honeycomb lattice. This is a special case of a bipartite random-hopping problem, belonging to the symmetry class BDI in the Altland-Zirnbauer classification.\cite{alzi97}
The single-particle properties of such systems have been analyzed using various techniques:\cite{gade93,ludwig94,motrunich02,mudry03,yamada04} All single-particle states at non-zero energies are exponentially localized, and the resulting density of states at low energies follows the form\cite{motrunich02,mudry03}
\begin{equation}
\label{motrudos}
\rho(\w) \propto \frac{1}{\w} \exp\left( -c|\ln\w|^{1/x}\right)
\end{equation}
with $x=3/2$. This immediately implies a corresponding singular behavior for the specific-heat coefficient $C/T$. However, the asymptotic form \eqref{motrudos} is only realized below an extremely small energy scale which depends on the disorder strength \cite{motrunich02} and is typically not accessible in numerical simulations.

In the application to the Kitaev model, two further aspects are important:
(i) For strong disorder, the ground state may not be located in the flux-free sector -- this will be discussed in Section~\ref{sec:trans}.
(ii) Even if the ground state is in the flux-free sector, the flux gap $\Delta$ may become small, and many-body states in excited flux sectors become important for temperatures $T\gtrsim\Delta$.


\section{Numerical results: Disordered system}
\label{sec:resdis}

\begin{figure}[t!]
  \centering
    \includegraphics[width=0.45\textwidth]{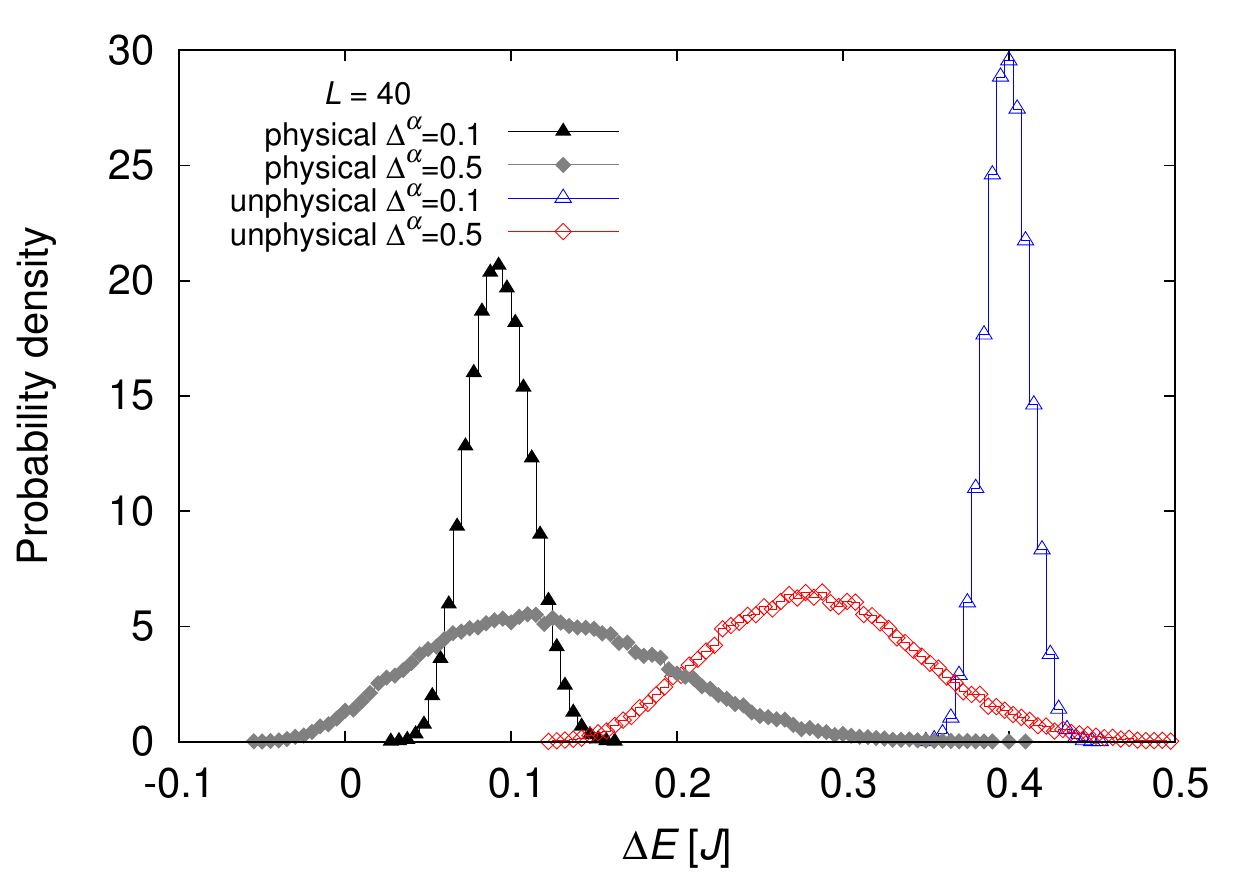}
    \includegraphics[width=0.45\textwidth]{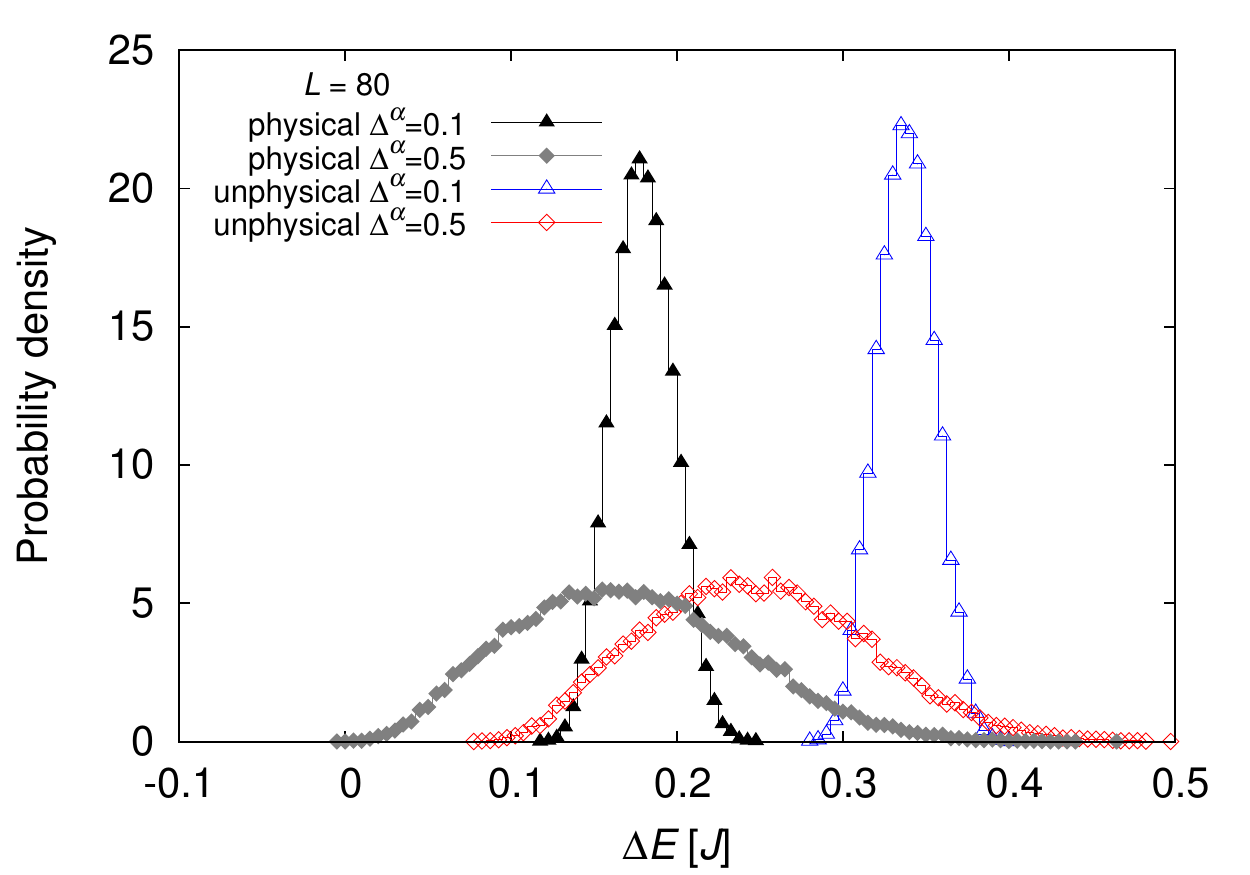}
\caption{
Distribution of the local flux gap for the isotropic Kitaev model with bond disorder, calculated for $L=40$ (top) and $L=80$ (bottom) and two different values of the disorder strength $\Delta^\alpha$. Shown are the results for both the ``physical'' (closed symbols) and the ``unphysical'' (open symbols) gap, calculated according to Eqs.~\eqref{fluxgap1} and \eqref{fluxgap2}, respectively.
}
\label{fig:gap80}
\end{figure}

\begin{figure}[t!]
  \centering
    \includegraphics[width=0.45\textwidth]{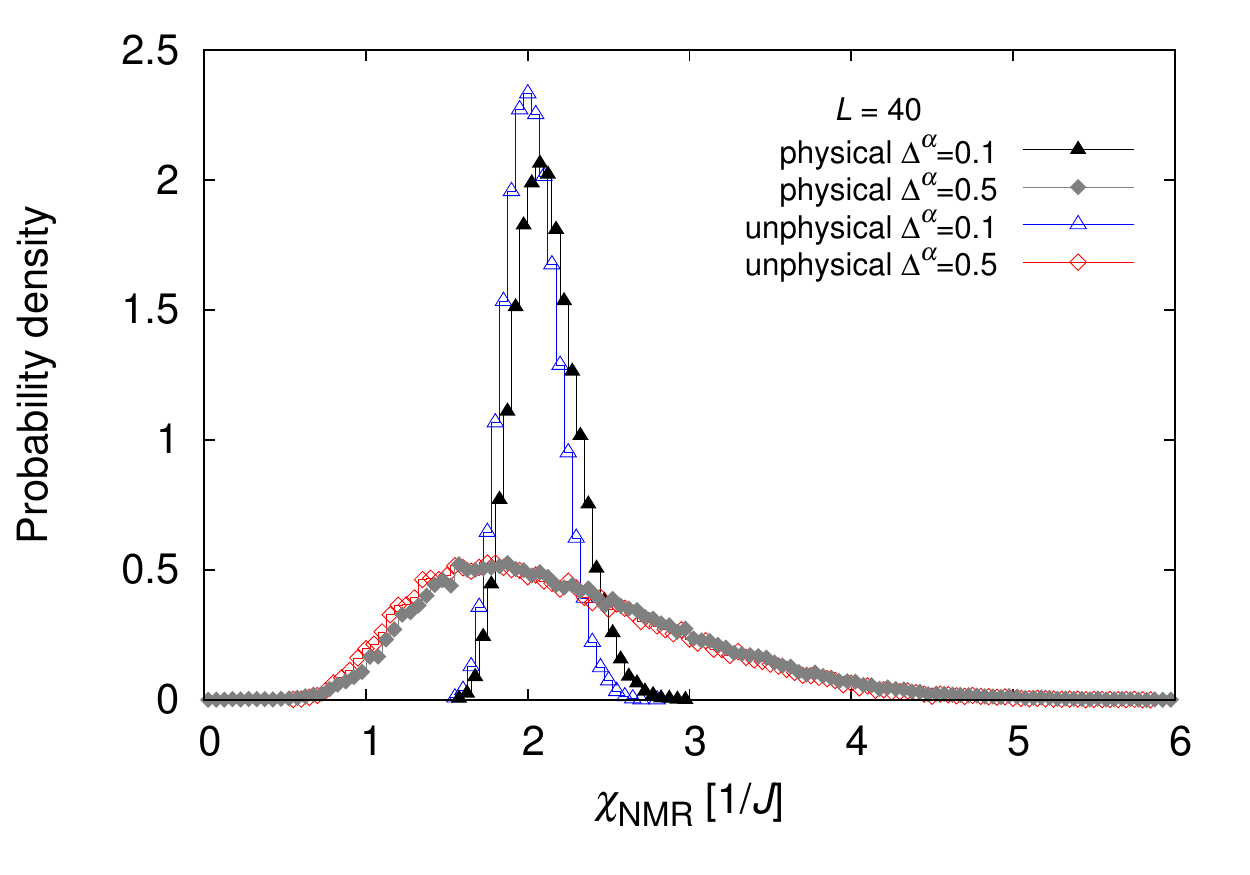}
    \includegraphics[width=0.45\textwidth]{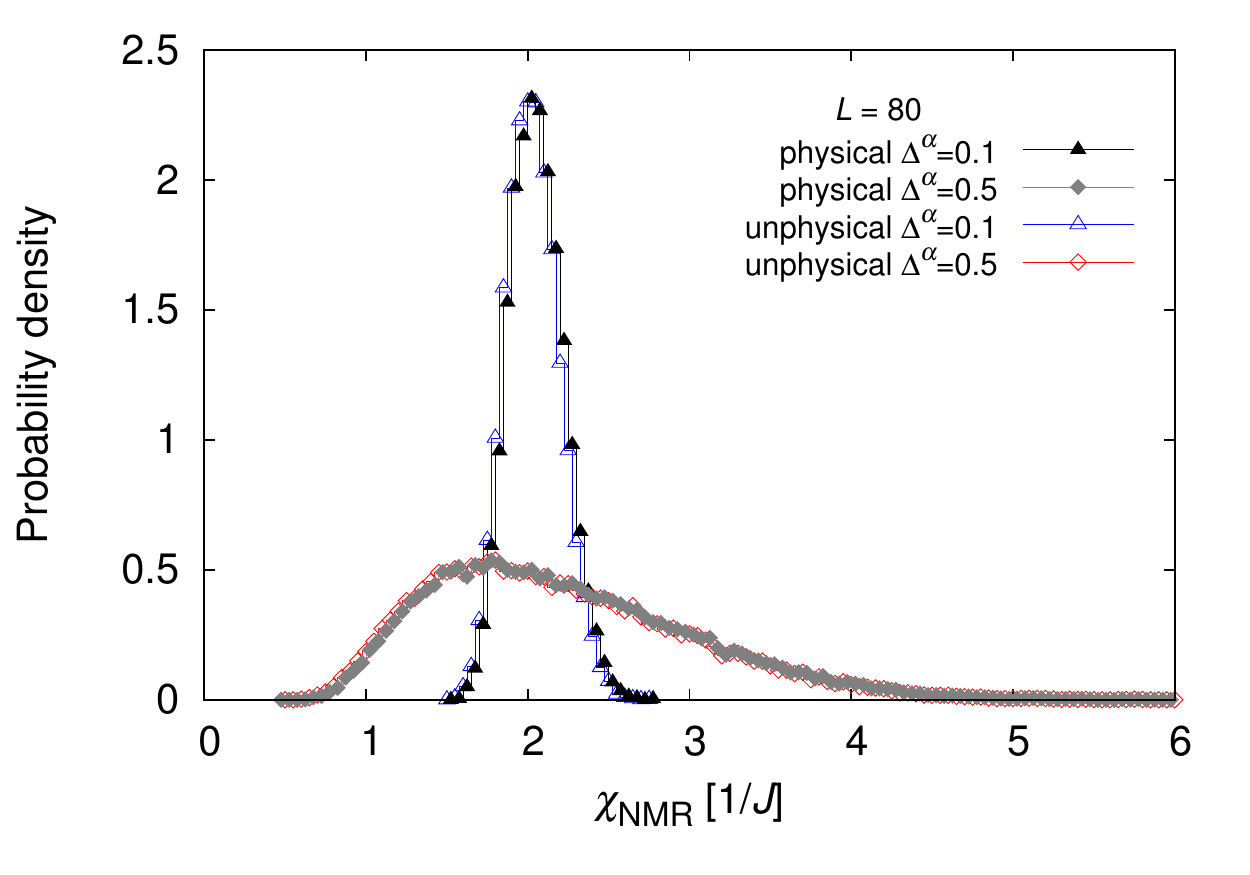}
\caption{
Distribution of the local (NMR) susceptibility, Eq.~\eqref{chinmr}, for the isotropic Kitaev model with bond disorder, calculated for $L=40$ (top) and $L=80$ (bottom) and two values of the disorder strength $\Delta^\alpha$. As before, ``physical'' (``unphysical'') represent the results obtained for one (zero) matter fermions in the flux-free ground state.
}
\label{fig:nmr}
\end{figure}

\subsection{Flux gap}

Figure~\ref{fig:gap80} shows histograms of the local flux gap, $\Delta E_{ij}$, for Kitaev models with bond disorder. This local gap is defined as in Eqs.~\eqref{fluxgap1} and \eqref{fluxgap2}, with the specific two-flux state obtained by flipping the $(ij)$ bond. We note that the selection rules for physical states continue to apply for the moderate disorder considered here.
Comparing the $L=40$ and $L=80$ data, strong finite-size effects are apparent which are inherited from the disorder-free situation, see the results in Fig.~\ref{fig:gap}. The following discussion thus mainly applies to the $L=80$ data.

For weak disorder, $\Delta^\alpha/J=0.1$, the gap distribution is essentially symmetric, with a relative width which roughly matches that of the coupling-constant distribution.
For strong disorder, $\Delta^\alpha/J=0.5$, the gap distribution widens and becomes slightly asymmetric. Its mean value is shifted downwards relative to the clean case (there $\Delta E_p/J=0.173$ and $\Delta E_u/J=0.345$ for $L=80$). Furthermore, cases with $\Delta E<0$ appear, i.e., the ground state is not in the flux-free sector. The significance of this finding will be discussed in Section~\ref{sec:trans}.

\subsection{Static susceptibility}

With an eye towards nuclear-magnetic-resonance experiments, we consider the local susceptibility
\begin{equation}
\label{chinmr}
\chi_\text{NMR}(i)=\sum_j \chi^{\alpha\alpha}_{ij}
\end{equation}
which is proportional to the resonance frequency in NMR experiments. We recall that, in the Kitaev model, $\chi_{ij}=0$ beyond nearest-neighbor distance, i.e, there are only on-site and nearest-neighbor contributions to $\chi_\text{NMR}$.

Results for the distribution of $\chi_\text{NMR}(i)$ are displayed in Fig.~\ref{fig:nmr}. While weak disorder again produces an essentially symmetric distribution with a relative width corresponding to that of the coupling-constant distribution, strong disorder produces a distinctly asymmetric shape with a tail at large values of $\chi$. The reason is in the strong fluctuations of the flux gap, Fig.~\ref{fig:gap80}, considering that $\chi \propto 1/\Delta E$. We note that in evaluating $\chi$ we have {\em assumed} the ground state to be flux-free, and consequently have discarded the rare events with $\Delta E<0$.

Interestingly, and in striking contrast to the results for the flux gap in Fig.~\ref{fig:gap80}, we find that the physical and unphysical results for the $\chi$ distribution are almost identical at $L=80$. The explanation is similar to that given in Section~\ref{sec:fssus}: Although the physical and unphysical cases have contributions to $\chi$ with rather different excitation energies, the corresponding matrix elements are small for large $L$. For instance, the zero-particle contribution to the physical susceptibility, with the excitation energy being the flux gap $\Delta E_p$ according to Eq.~\eqref{fluxgap1}, has a weight scaling as $N^{-1}$.

\subsection{Dynamic susceptibility}

\begin{figure}[t!]
  \centering
    \includegraphics[width=0.45\textwidth]{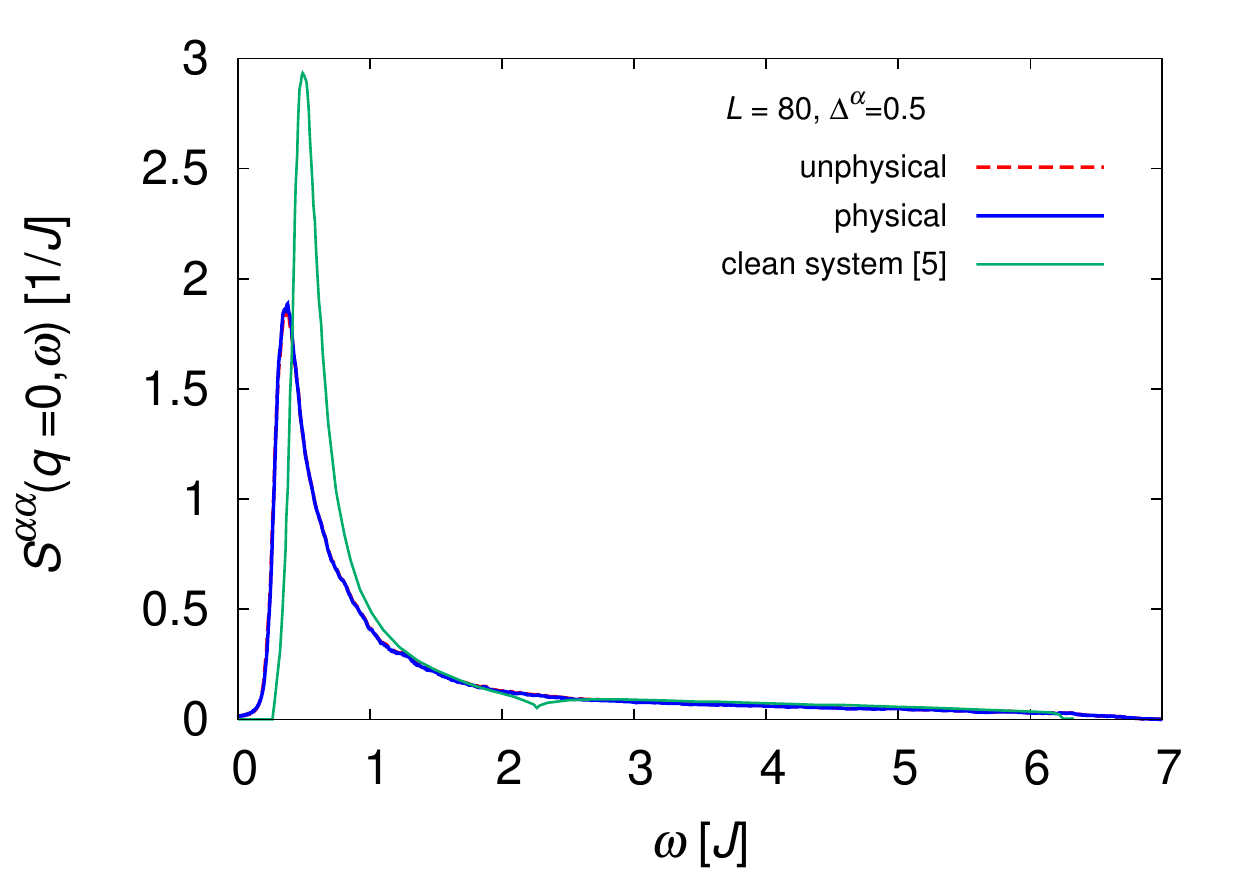}
\caption{
Dynamic structure factor as in Fig.~\ref{fig:s40}, but now for the Kitaev model of size $L_1=L_2=80$ with box-type bond disorder of strength $\Delta^\alpha/J=0.5$. The artificial broadening is smaller than in Fig.~\ref{fig:s40}: $\delta/J=0.01$. The clean-system result\cite{knolle} is shown for comparison.
}
\label{fig:diss80}
\end{figure}

As a further example, we plot the dynamic structure factor in the presence of bond disorder in Fig.~\ref{fig:diss80}. As disorder tends to smear the flux gap, the gap in the structure factor is filled. This is accompanied by a shift of weight to lower energies, as expected from Fig.~\ref{fig:gap80}. Disorder-induced changes at higher energies are minimal.
Consistent with the above discussion, there is essentially no difference between the physical and unphysical results at $L=80$ in Fig.~\ref{fig:diss80}.

\section{Transition out of flux-free state}
\label{sec:trans}

Our numerical results show that, with increasing bond disorder, the ground state of a {\em finite-size} Kitaev model is no longer located in the flux-free sector. Instead, the ground state displays a finite flux density, where fluxes occur in the system at random positions which depend on the disorder realization. We note that such a state is trivially realized for box disorder with $\Delta^\alpha/J>1$, as this implies the existence of bonds with flipped sign which can be compensated by placing flux pairs adjacent to these bonds (equivalent to choosing $u=-1$ on the respective bonds). More interesting is the possible occurrence of such a random-flux state for $\Delta^\alpha/J<1$ where all bond strengths are positive.
Notably, the numerics also indicates that the tendency towards ground-state fluxes diminishes with increasing $L$ (see e.g. Fig.~\ref{fig:gap80}), such that definite conclusions about the thermodynamic limit cannot be drawn.

\begin{figure}[t!]
  \centering
    \includegraphics[width=0.45\textwidth]{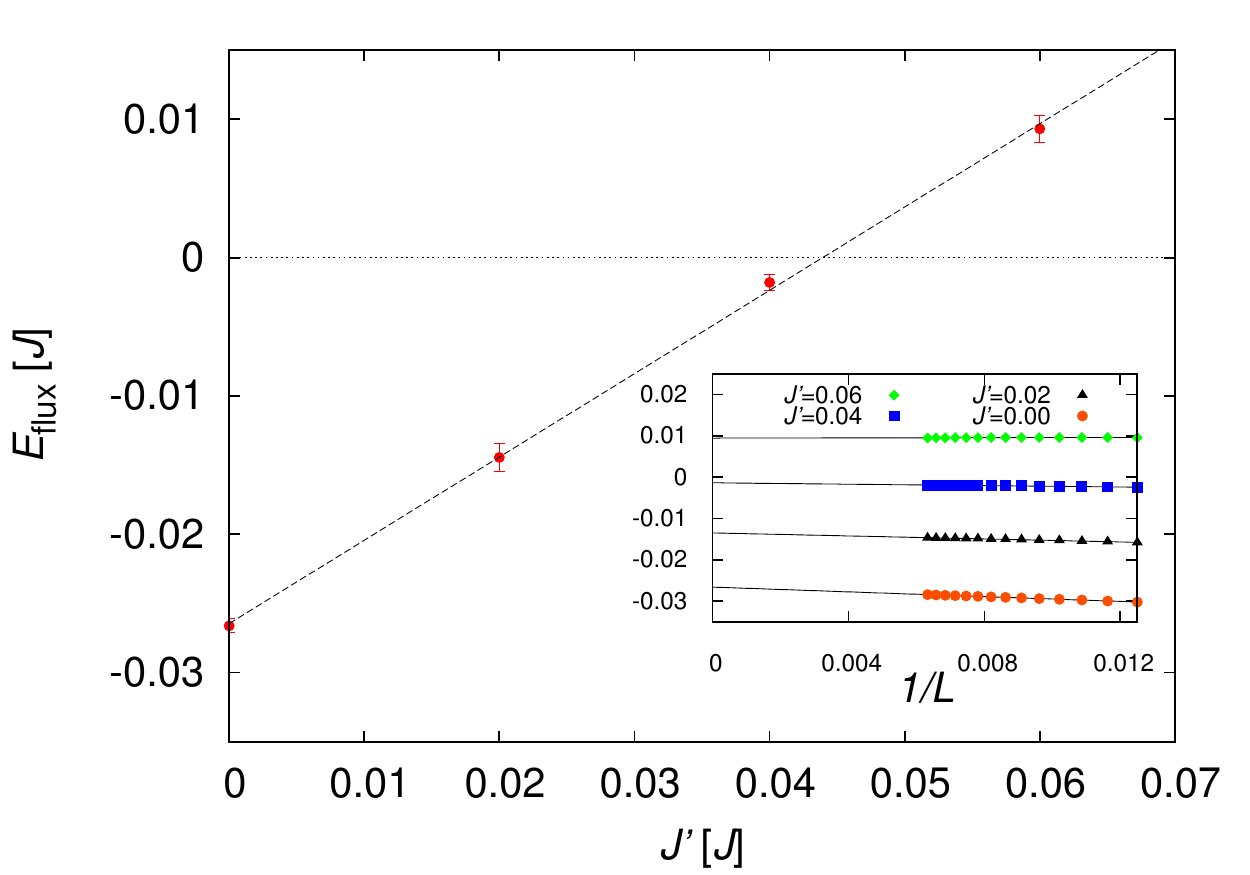}
\caption{
Flux energy $E_{\rm flux}(J')$ for an isotropic Kitaev model with a single defect site which has three weak bonds of strength $J'$ to its neighbors. $E_{\rm flux}<0$ implies that the defect binds a flux; $E_{\rm flux}(0)/J=-0.027$ is the known result for a vacancy from Ref.~\onlinecite{willans10}. $E_{\rm flux}(J')$ has been calculated as the energy difference between a two-flux state, with one flux in one of the three defect plaquettes and one flux at maximum distance away from the defect, and the flux-free state, and the energy of an isolated flux (for the same $L$) has been subtracted. The inset shows the finite-size scaling of $E_{\rm flux}$, the error bars in the main panel arise from uncertainties in the $L\to\infty$ extrapolation. The dashed line is a linear fit.
}
\label{fig:fluxbind}
\end{figure}

However, we are able to provide a {\em general} argument in favor of a non-trivial transition to a random-flux state which applies to the thermodynamic limit.
A key ingredient is the observation of Refs.~\onlinecite{willans10,willans11} that a vacancy site gains a finite amount of energy by binding a flux. Consider now the more general situation where a single defect site is surrounded by three bonds of strength $J'$ and embedded in an otherwise homogeneous Kitaev model with couplings $J$. While $J'=0$ corresponds to the vacancy case, this site will also bind a flux for finite small $J'$. This is shown in Fig.~\ref{fig:fluxbind}: For all $0<J'<J_{\rm min}$ with $J_{\rm min}/J\approx 0.04$ the energy of the state with a flux bound in one of the three plaquettes adjacent to the defect is lower than that of the flux-free state. More generally, a defect site surrounded by three bonds of a strength in the interval $[0,J_{\rm min}]$ will bind a flux.

Now, for box disorder with strength $\Delta^\alpha$ the minimum coupling strength is $J-\Delta^\alpha$, thus that for any $\bar{J}>J-\Delta^\alpha$ there is a finite probability to find local configurations which have (i) three bonds emanating from one site with strength smaller than $\bar{J}$ and (ii) all surrounding bond strengths arbitrarily close to $J$.
This is exactly the condition for locally binding a flux, provided that $\bar{J}<J_{\rm min}$. We conclude that a random-flux state must be realized for disorder strengths with $\Delta^\alpha>J-J_{\rm min}$. This proves the existence of a transition -- from zero flux to random flux -- somewhere in the interval $0<\Delta^\alpha/J<1-J_{\rm min}/J\approx 0.96$.


\section{Summary}

Our study of Kitaev's honeycomb model with bond disorder has lead to twofold results:
On the one hand, we have dealt with the selection of physical states in the Majorana representation. Extending earlier work, we have shown that the ground state of the gapless Kitaev model with periodic boundary conditions generically contains one matter fermion excitation. This causes significant finite-size effects for observables, as illustrated for the flux gap. We have also discussed the difference in state selection between the cases with periodic and open boundary conditions. Obviously, this state selection is of relevance for all numerical studies of Kitaev models using Majorana fermions. It will be interesting to extend this analysis to other tricoordinated lattices where the Kitaev model can also be solved exactly; work in this direction is in progress.

On the other hand, we have numerically determined the static and dynamic spin susceptibility in the presence of bond disorder. In particular, we have calculated the distribution of local susceptibilities which determines the NMR lineshape. For large disorder, we predicted a transition to a random-flux state. A detailed study of this transition is left for future work.


\acknowledgments

We thank W. Brenig, D. Kovrizhin, and, in particular, J. Chalker, J. Knolle, T. Meng, and R. Moessner
for discussions.
This research was supported by the DFG through SFB 1143 and GRK 1621 as well as by the Helmholtz association through VI-521. FZ also acknowledges support by the Stiftung der Deutschen Wirtschaft and by the International Max Planck Research School on Dynamical Processes in Atoms, Molecules, and Solids.


\appendix

\section{Parity of matter fermion excitations}
\label{sec:physicalgapless}

\subsection{Gapless phase}

The purpose of this appendix is to prove that all flux-free physical states in the gapless phase of a translation-invariant Kitaev model with periodic boundary conditions contain an {\em odd} number of $\hat{a}_m$ fermion excitations. This supersedes the results of Ref.~\onlinecite{loss}, but is consistent with their Fig.~3.

The proof is based on insights from Ref.~\onlinecite{loss} which we lay out first. The flux-free sector is characterized by all $u_{ij}=1$. Then the eigenmodes of $\HM$ are diagonal in momentum space:\cite{kitaev06}
\begin{equation}
\label{momentum}
\HM = \sum_{{\bf q}}\left|f({\bf q})\right|(2\hat{a}^{\dagger}_{{\bf q}}\hat{a}_{{\bf q}}-1)
\end{equation}
with $f({\bf q})=J^x e^{i{\bf q \cdot e}_1}+J^y e^{i{\bf q \cdot e}_2}+J^z$. The spectrum
$\left| f({\bf q})\right|$ is gapped if $J^z > J^x + J^y$ or permutations, and gapless otherwise.
The reciprocal lattice is defined by the vectors ${\bf b}_{1,2}$, see Fig.~\ref{fig:BC}. For any finite lattice the Brillouin zone is reduced to a finite set of wavevectors ${\bf q}$, which can be partitioned into three sets $\Omega$ and $\Omega\pm$. We assign ${\bf q} \in \Omega $ if $\pm {\bf q}$ are equivalent (up to reciprocal lattice vectors); there are at most four wavevectors in $\Omega$, namely ${\bf 0},{\bf b}_1/2, {\bf b}_2/2$, and $({\bf b}_1+{\bf b}_2)/2$. The remaining $\bf q$ are partitioned such that $\pm \bf q$ belong to two distinct sets $\Omega_\pm$.
One can then derive the explicit formula for the determinant of the transformation matrix \cite{loss}
\begin{equation}
\det (Q^u)=-1^{\gamma+N^2},
\end{equation}
valid for the flux-free sector, where $N=L_1 L_2$, and $\gamma$ is the number of reciprocal vectors
${\bf q} \in \Omega $ with $f({\bf q}) < 0$. Together with the geometric factor \eqref{eq:gf},
we can now rewrite
\begin{equation}
\label{mudef}
(-1)^\theta \det (Q^u)=(-1)^{\gamma+L_1+L_2+L_1^2L_2^2+L_1M-M^2}\equiv(-1)^\mu\,.
\end{equation}

Although $\gamma$ depends in a non-trivial way on the boundary conditions $L_{1,2}$ and $M$ as well as on the couplings $J_{x,y,z}$, we can calculate it for any given choice of $L_{1,2}$, $M$.
Since ${\bf b}_i{\bf e}_j=2\pi \delta_{ij}$, it is easy to see that in the gapless phase only
$f\left( \frac{{\bf b}_1+{\bf b}_2}{2}\right)=J^z-J^x-J^y$ is less then $0$. Therefore $\gamma=1$ if
$({\bf b}_1+{\bf b}_2)/2 \in \Omega$ and $\gamma = 0$ otherwise.
The allowed $\bf q$ vectors are determined by the conditions
\begin{align}
    e^{i{\bf q}L_1{\bf b}_1} &= 1,
\\  e^{i{\bf q}(L_2{\bf b}_2+M{\bf b}_1)} &= 1,
\end{align}
with ${\bf q}=q_1{\bf b}_1+q_2{\bf b}_2$. Therefore $\gamma=1$ if $L_1 = 2n_1$ and $L_2+M = 2n_2$ ($n_{1,2}\in Z$).
Enumerating all eight combinations of parities of $L_{1,2}$ and $M$ yields the results in table~\ref{tab:eo}, showing that $-1^\mu = -1$ in all cases.
Using equation \eqref{Dmatter} this implies that, in the flux-free case where $N_\chi=0$, the physical Majorana states must have an odd number of matter fermion excitations, $\pi= (-1)^{N_a}\overset{!}{=}-1$.

\begin{table}[t]
  \centering
    \begin{tabular}{|ccc|c|cc||c|} \hline

      $L_1$     & $L_2$   & $M$ & $\gamma$  &  $(L_1 L_2)^2$     & $L_1M$      & $(-1)^\mu$ \\  \hline

    +         & +        & +    & 1   &  +            & +          & $-1$ \\
    +         & $-$      & $-$  & 1   &  +            & +          & $-1$ \\
              &          &      &       &               &            &    \\
    +         & +        & $-$  & 0     &  +            & +          & $-1$ \\
    +         & $-$      & +    & 0     &  +            & +          & $-1$ \\
    $-$       & +        & +    & 0     &  +            & +          & $-1$ \\
    $-$       & $-$      & $-$  & 0     &  $-$             & $-$           & $-1$ \\
    $-$       & +        & $-$  & 0     &  +            & $-$           & $-1$ \\
    $-$       & $-$      & $+$  & 0     &  $-$             & $+$           & $-1$ \\
\hline
    \end{tabular}%
    \caption{This table shows $(-1)^\theta \det (Q^u) \equiv (-1)^\mu$ for the gapless phase in relation to the boundary conditions $L_{1,2},M$ (where $+$ and $-$ refer to even and odd values, respectively) and the resulting $\gamma$, see text.}
  \label{tab:eo}%
\end{table}

From this result one can further deduce that $N_{a}$ for states in the two-flux sector, at and near the isotropic point, is even. This flux sector has $N_{\chi}=1$, and Eq.~\eqref{Dmatter} $(-1)^\Theta \det(Q^u)(-1)^{N_{\chi}}(-1)^{N_{a}} \overset{!}{=} 1$ implies that $N_{a}$ must be even as long as the signs of $\det(Q^u)$ in the zero-flux and two-flux sectors are identical. The latter applies near the isotropic point, but not in the entire gapless phase.\cite{knolle}


\begin{figure}[t]
  \centering
    \includegraphics[width=0.43\textwidth]{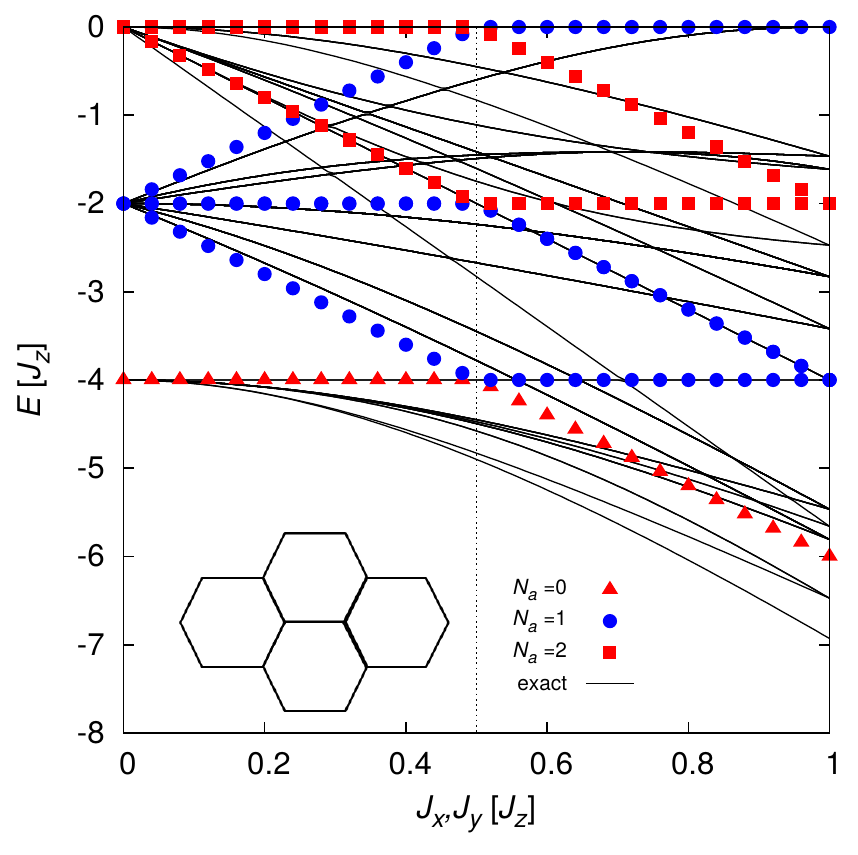}
\caption{
Lower half of the many-body spectrum of an anisotropic $2\times2$ Kitaev model with $J^x=J^y \leq J^z$ as function of $J^x/J^z$, with the system geometry shown in the inset.
Lines: Eigenenergies obtained by exact diagonalization of the spin Hamiltonian.
Symbols: Eigenenergies of the Majorana Hamiltonian in the flux-free sector, $u_{ij}=1$.
$N_a$ is the number of matter fermion excitations.
At (and near) the isotropic point, $N_a=0,2$ states are unphysical (red) while $N_a=1$ states are physical (blue).
The vertical dashed line indicates the boundary between the gapped and gapless phases.\cite{kitaev06}
} \label{fig:E0}
\end{figure}

\begin{figure}[t]
  \centering
    \includegraphics[width=0.43\textwidth]{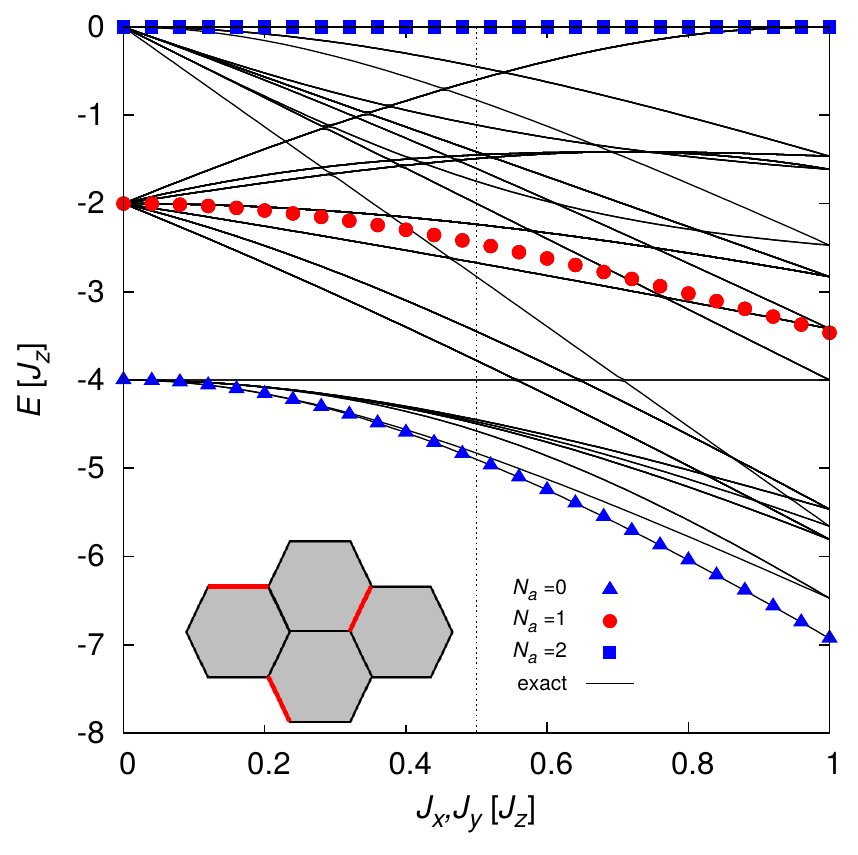}
\caption{
Same as Fig.~\ref{fig:E0}, but for the four-flux sector with $W_1 = W_2=-1$. The bonds with
$u_{ij}=-1$ are shown in light (red) color in the inset.
Here, $N_a=0,2$ states are physical (blue) while $N_a=1$ states are unphysical (red) near the isotropic point.
} \label{fig:E4}
\end{figure}

\begin{figure}[!b]
  \centering
    \includegraphics[width=0.43\textwidth]{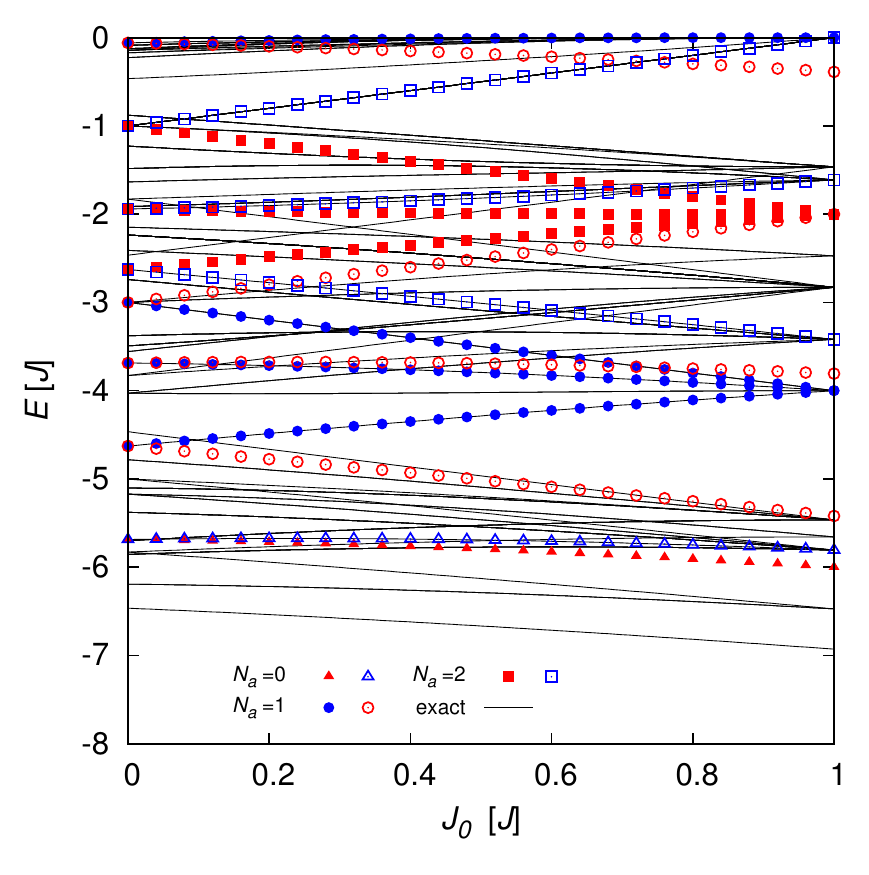}
\caption{
Same as Fig.~\ref{fig:E0}, but now for an isotropic model where a single bond has a different exchange strength $J_0 \neq J$.
Full (open) symbols correspond to the Majorana eigenenergies in the sectors with zero flux (two fluxes, with a flux pair adjacent to the $J_0$ bond), respectively. As before, blue (red) symbols denote physical (unphysical) states.
} \label{fig:single}
\end{figure}

\subsection{Gapped phase}

Although a similar analysis may be performed for the gapped phase of the Kitaev model, it turns out that the different parity combinations of $L_{1,2}$ and $M$ come with different signs for $(-1)^{\mu}$. In particular, the dependence $\gamma(L_1,L_2,M)$ is different from that in the gapless phase, and $\gamma$ can now take all values from $0$ to $3$. As a result, a unique conclusion similar to the gapless phase cannot be reached. Moreover, the small flux gap in combination with the large fermionic gap can lead to the physical ground state having excited flux pairs but no fermions, see also Fig.~5 of Ref.~\onlinecite{loss}.


\section{Spectrum for $L_1=L_2=2$}
\label{app:exact}

In this appendix we verify the analysis in Section~\ref{sec:phys} by comparing the eigenenergies of $\HK$, obtained by exact diagonalization of the spin Hamiltonian, with the energies of the many-body Majorana states, both physical and unphysical.

We choose a small system with $L_1=L_2=2$ and $M=0$. Here the Dirac point does not belong to the discrete partitioning of the Brillouin zone, such that all excitation energies of matter fermions, $\e_m$, are non-zero.

\subsection{Periodic boundary conditions and varying anisotropy}

To illustrate the unphysical character of the zero-flux fermion-free state, we show in Fig.~\ref{fig:E0} the many-body Majorana energies in the zero-flux sector, together with all $2^8=256$ eigenenergies of the spin Hamiltonian, for varying spin anisotropy.

In the entire gapless phase, $1/2 \leq J^{x,y}/J^z \leq 1$, the Majorana states with even number $N_a$ of matter fermion excitations do not correspond to any of the physical states, whereas the Majorana states with odd $N_a$ match the physical spectrum. Interestingly, this behavior is reversed in the gapped phase, $0\leq J^{x,y}/J^z < 1/2$, where now the states with even $N_a$ are physical.

We have repeated this analysis in all flux sectors. As an example, we show the flux sector containing the ground state, here with fluxes through all plaquettes, in Fig.~\ref{fig:E4}. The physical states in this sector have an even number of excited matter fermions in both phases.

Interestingly, in the three flux sectors without plaquette fluxes but with a flux through
at least one of the torus holes, i.e., $W_1=-1$, $W_2=1$, $W_1=1$, $W_2=-1$, and $W_1=W_2=-1$, the even-$N_a$ states are found to be physical.

\subsection{Varying a single bond}

To underline the arguments concerning missing bonds and open boundary conditions in Section~\ref{sec:openbc} we now consider an isotropic $L_1=L_2=2$ system where we vary the exchange strength $J_0$ on one bond keeping the other couplings fixed at $J$. Fig.~\ref{fig:single} shows the Majorana energies both in the zero-flux and two-flux sectors, in the latter case with the flux pair located adjacent to the $J_0$ bond, together with the exact spectrum.

For any non-zero $J_0$, the states with odd (even) $N_a$ are physical in the zero-flux (two-flux) sector, respectively, consistent with our reasoning above. However, for $J_0=0$, all matter Majorana states become physical: This is a consequence of the zero mode constructed from gauge Majorana fermions in the presence of a missing bond, see Section~\ref{sec:openbc}. Consistent with this, the energy difference between the zero-flux and two-flux states vanishes as the flux pair has no observable impact if it surrounds the $J_0=0$ bond.



\begin{thebibliography}{99}

\bibitem{balents10}
L. Balents,
Nature {\bf 464}, 199 (2010).

\bibitem{kitaev06}
A.~Kitaev, Ann. Phys. (N.Y.) {\bf 321}, 2 (2006).

\bibitem{compass_rmp}
For a comprehensive review on compass and Kitaev models see:
Z. Nussinov and J. van den Brink,
Rev. Mod. Phys. {\bf 87}, 1 (2015) and preprint arXiv:1303.5922.

\bibitem{shank}
G. Baskaran, S. Mandal, and R. Shankar,
Phys. Rev. Lett. {\bf 98}, 247201 (2007).

\bibitem{knolle}
J. Knolle, D. L. Kovrizhin, J.~T. Chalker, and R. Moessner,
Phys. Rev. Lett. {\bf 112}, 207203 (2014).

\bibitem{willans10}
A. J. Willans, J. T. Chalker, and R. Moessner,
\prl {\bf 104}, 237203 (2010).

\bibitem{willans11}
A. J. Willans, J. T. Chalker, and R. Moessner,
\prb {\bf 84}, 115146 (2011).

\bibitem{dhochak10}
K. Dhochak, R. Shankar, and V. Tripathi,
Phys. Rev. Lett. {\bf 105}, 117201 (2010).

\bibitem{yao07}
H. Yao and S. A. Kivelson,
Phys. Rev. Lett. {\bf 99}, 247203 (2007).

\bibitem{kiv1}
H. Yao, S.-C. Zhang, and S. A. Kivelson,
Phys. Rev. Lett. {\bf 102}, 217202 (2009).

\bibitem{baskaran09}
G. Baskaran, G. Santhosh, and R. Shankar,
preprint arXiv:0908.1614

\bibitem{kamfor10}
M. Kamfor, S. Dusuel, J. Vidal, and K. P. Schmidt,
J. Stat. Mech. P08010 (2010).

\bibitem{kells11}
G. Kells, J. Kailasvuori, J. K. Slingerland, and J. Vala,
New J. Phys. {\bf 13}, 095014 (2011).

\bibitem{siyu08}
T. Si and Y. Yu, Nucl. Phys. B {\bf 803}, 428 (2008).

\bibitem{mandal09}
S. Mandal and N. Surendran,
Phys. Rev. B {\bf 79}, 024426 (2009).

\bibitem{trebst14}
M. Hermanns and S. Trebst,
Phys. Rev. B {\bf 89}, 235102 (2014).

\bibitem{loss}
F. L. Pedrocchi, S. Chesi, and D. Loss,
\prb {\bf 84}, 165414 (2011).

\bibitem{Cha10}
J.~Chaloupka, G.~Jackeli, and G.~Khaliullin,
Phys. Rev. Lett. {\bf 105}, 027204 (2010).

\bibitem{Jia11}
H.-C.~Jiang, Z.-C.~Gu, X.-L.~Qi, and S.~Trebst,
Phys. Rev. B {\bf 83}, 245104 (2011).

\bibitem{Reu11}
J.~Reuther, R.~Thomale, and S.~Trebst,
Phys. Rev. B {\bf 84}, 100406 (2011).

\bibitem{Bha12}
S.~Bhattacharjee, S.-S.~Lee, and Y.B.~Kim,
New J. Phys. {\bf 14}, 073015 (2012).

\bibitem{Cha13}
J.~Chaloupka, G.~Jackeli, and G.~Khaliullin,
Phys. Rev. Lett. {\bf 110}, 097204 (2013).

\bibitem{perkins12}
C. Price and N. B. Perkins,
Phys. Rev. Lett. {\bf 109}, 187201 (2012);
Phys. Rev. B {\bf 88}, 024410 (2013).

\bibitem{eamv14}
E. C. Andrade and M. Vojta,
Phys. Rev. B {\bf 90}, 205112 (2014).

\bibitem{Sin10} Y.~Singh and P.~Gegenwart,
Phys. Rev. B {\bf 82}, 064412 (2010).

\bibitem{Liu11}
X.~Liu, T.~Berlijn, W.-G.~Yin, W.~Ku, A.~Tsvelik, Y.-J.~Kim, H.~Gretarsson, Y.~Singh, P.~Gegenwart, and J.P.~Hill,
Phys. Rev. B {\bf 83}, 220403 (2011).

\bibitem{Sin12}
Y.~Singh, S.~Manni, J.~Reuther, T.~Berlijn, R.~Thomale, W.~Ku, S.~Trebst, and P.~Gegenwart,
Phys. Rev. Lett. {\bf 108}, 127203 (2012).

\bibitem{Choi12}
S. K.~Choi, R.~Coldea, A. N.~Kolmogorov, T.~Lancaster, I. I.~Mazin, S. J.~Blundell, P. G.~Radaelli, Y.~Singh, P.~Gegenwart, K. R.~Choi, S.-W.~Cheong, P. J.~Baker, C.~Stock, and J.~Taylor,
Phys. Rev. Lett. {\bf 108}, 127204 (2012).

\bibitem{Maz12}
I. I.~Mazin, H. O.~Jeschke, K.~Foyevtsova, R.~Valenti, and D. I.~Khomskii,
Phys. Rev. Lett. {\bf 109}, 197201 (2012).

\bibitem{kee14}
J. G. Rau, E. K.-H. Lee, and H.-Y. Kee,
Phys. Rev. Lett. {\bf 112}, 077204 (2014).

\bibitem{rachel14}
J. Reuther, R. Thomale, and S. Rachel,
Phys. Rev. B {\bf 90}, 100405(R) (2014).

\bibitem{perkins14}
Y. Sizyuk, C. Price, P. W\"olfle, and N. B. Perkins,
Phys. Rev. B {\bf 90}, 155126 (2014).

\bibitem{kimchi14}
I. Kimchi, R. Coldea, and A. Vishwanath,
preprint arXiv:1408.3640.

\bibitem{lahtinen14}
V. Lahtinen, A. W. W. Ludwig, and S. Trebst,
Phys. Rev. B {\bf 89}, 085121 (2014).

\bibitem{fiete11}
V. Chua and G. A. Fiete,
\prb {\bf 84}, 195129 (2011).


\bibitem{halasz}
G.~B. Hal{\'a}sz, J.~T. Chalker, and R. Moessner,
Phys. Rev. B {\bf 90}, 035145 (2014).

\bibitem{willansfoot}
Ref.~\onlinecite{willans11} quotes the equation $D = (-1)^{N_f} (-1)^{N_\chi}$ which apparently misses the factor $(-1)^\theta$ from Eq.~\eqref{dbond}.

\bibitem{BlaRip}
J. P. Blaizot and G. Ripka, \textit{Quantum Theory Of Finite Systems},
MIT Press (1985).

\bibitem{kn_raman}
The expression for the matrix element in Eq.~\eqref{tpart} has been independently derived in:
J. Knolle, G.-W. Chern, D. L. Kovrizhin, R. Moessner, and N. B. Perkins,
Phys. Rev. Lett. {\bf 113}, 187201 (2014).

\bibitem{lieb}
E. H. Lieb,
\prl {\bf 73}, 2158 (1994).


\bibitem{alzi97}
A. Altland and M. R. Zirnbauer,
\prb {\bf 55}, 1142 (1997).

\bibitem{gade93}
R. Gade and F. Wegner, Nucl. Phys. B {\bf 360}, 213 (1991);
R. Gade, {\em ibid.} {\bf 398}, 499 (1993).

\bibitem{ludwig94}
A. W. W. Ludwig, M. P. A. Fisher, R. Shankar, and G. Grinstein,
\prb {\bf 50}, 7526 (1994).

\bibitem{motrunich02}
O. Motrunich, K. Damle, and D. A. Huse,
\prb {\bf 65}, 064206 (2002).

\bibitem{mudry03}
C. Mudry, S. Ryu, and A. Furusaki,
\prb {\bf 67}, 064202 (2003).

\bibitem{yamada04}
H. Yamada and T. Fukuhi,
Nucl. Phys. B {\bf 679}, 632 (2004).


\bibitem{orth_foot}
For system sizes with $L \mod 3=0$ we observe that $|\det X|=|\langle\mz|\lambda_0 \rangle|^2$ becomes anomalously small, presumably due to the degeneracy of the lowest-energy zero-flux state due to the presence of a zero-energy matter fermion. $|\det X|$ being small causes large finite-size effects in $S(\w)$.

\bibitem{knollefoot}
Ref.~\onlinecite{knolle} did not take into account the parity condition for the matter fermion excitiations, but their results are correct in the thermodynamic limit.


\end{thebibliography}
\end{document}